\documentclass[%
 aip,
 JAP,%
 amsmath,amssymb,
 preprint,%
]{revtex4-1}

\usepackage{graphicx}
\usepackage{dcolumn}
\usepackage{bm}
\usepackage{color,soul}
\usepackage{amsmath}
\usepackage{textcomp}

\usepackage{subcaption}
\usepackage{amsmath}
\usepackage{comment}
\def\x{{\bf{x}}}

\def\xo{\bf{x_0}}
\def\n{{ \bf n}}

\def\v{{ \bf v}}
\def \G{{G(\x,\xo)}}
\def \del{\partial}

\def\bnabla{{ \bf \nabla}}

\def\n{{\bf n}}

\def\l[{{[\![}}
\def\rt]{{]\!]}}

\begin{document}

\preprint{AIP/123-QED}

\title{Influence of the trap potential waveform on surface oscillation and breakup of a levitated charged drop}


\author{Mohit Singh}
\author{Neha Gawande}
\author{Rochish Thaokar}
\email{rochish@che.iitb.ac.in}
\affiliation{Department of Chemical Engineering, Indian  Institute of  Technology Bombay, Mumbai-400076.}

\date{\today}

\begin{abstract}
A charged droplet can be electrodynamically levitated in air using a quadrupole trap by typically applying sinusoidal electric field. When a charged drop is levitated it exhibits surface oscillations simultaneously building charge density due to continuous evaporation and subsequently undergoes breakup due to Rayleigh instability. In this work, we examined large-amplitude surface oscillations of a sub-Rayleigh charged drop and its subsequent breakup, levitated by various applied signals such as sine, square and ramp waveform at various imposed frequencies, using high-speed imaging (recorded at 100-130 thousand Frames Per Second (fps)). It is observed that the drop surface oscillates in sphere-prolate-sphere-oblate (SPSO) mode and seldom in the sphere-prolate-sphere (SPS) mode depending on the intricate interplay of various forces due to charge(q), the intensity of applied field ($\Lambda$) and shift of the droplet from the geometric center of the trap ($z_{shift}$). The Fast Fourier Transformation (FFT) analysis shows that the droplet oscillates with the forced frequency irrespective of the type of the applied waveform. While in the sinusoidal case, the nonlinearities are significant, in the square and ramp potentials, there is an admittance of all the harmonic frequencies of the applied potential. Interestingly, the breakup characteristics of a critically charged droplet is found to be unaffected by the type of the applied waveform. The experimental observations are validated with an analytical theory as well as with the Boundary Integral (BI) simulations in the potential flow limit and the results are found to be in a reasonable agreement.     
\end{abstract}

\pacs{Valid PACS appear here}
\maketitle
\section{Introduction}
The understanding of how inherently charged droplets oscillate, deform and break is vital in many technological applications such as ink-jet printing \cite{choi2008drop}, fuel injection \cite{shrimpton1999}, crop spraying \cite{bailey1986} and electrospray atomization for ion mass spectroscopy of biomolecules \cite{fenn1989}. Also, oscillating droplets or bubbles have been recognized an innovative experimental platform for understanding a wide variety of natural, analytical and biological applications such as interfacial reactions and rheology, thin-film, biosensors, and biophysical simulations. The experimental techniques involving oscillating droplet/bubble have several advantages such as a high rate of surface reaction and more substantial interfacial activity due to high surface-area-to-volume ratio, reduced consumption of valuable chemicals, and a well-defined environmental control due to miniaturized droplet platform. The controlled oscillations of a surface can be recognized as a basis for many analytical studies. For example, the sinusoidal oscillations in the presence of constant amplitude and imposed frequency of an applied force can be used to measure the surface dilational modulus of a molecular (surfactants or macromolecules) monolayer by estimating the time-varying interfacial tension.

Numerous methods such as pendent drop\cite{yeung2006, mashayek1998}, sessile drops \cite{von2003,ravera2010,chang2015}, acoustic levitation \cite{marston1980,shen2010}, electromagnetic levitation \cite{hill2012} and electrodynamic levitation \cite{singh2018surface} have been practised to investigate droplet oscillations. For understanding the oscillation response of a small-sized ($\sim$ 10-300$\mu$m) charged droplet, the electrodynamic levitation is considered to be one of the most efficient techniques. One of the potential reasons is that as the droplet size is small the gravity plays a negligible role in the steady-state droplet deformation and the droplet shape remains nearly symmetric throughout oscillations.

The capillary oscillations of an electrodynamically levitated charged droplet could be considered as an idealized system because the droplet gets levitated in a contact-free environment, thereby removing any solid boundary effect on the droplet oscillation characteristics. Thus, suitable modeling and development of analytical theory for capillary oscillations of a charged droplet in a quadrupole trap, together with relevant experiments, can make this system a reliable and accurate tool for characterizing the fluid properties such as surface tension and viscosity.

Towards this, many analytical studies \cite{feng1997, hasse1975, tsamopoulos1985} on the droplet oscillations have been reported. Very recently \citet{singh2018surface} reported theoretical analysis, in the small electro capillary limit, of surface oscillations of a charged liquid droplet levitated in an electrodynamic balance in the presence of sinusoidal applied waveform. The oscillations were ascertained to be governed by a potential flow description with viscous corrections. The effect of viscosity was exercised into consideration by including it in the normal stress balance condition of a potential flow limit. The analytical model was validated through numerous experiments involving the recording of surface oscillation dynamics at a very high frame rate of $\sim$ 100-130 thousand fps. An interesting question one can ask here is: what will be the effect of non-sinusoidal waveforms such as square and ramp on the characteristics of charged droplet oscillations.

The droplet levitation using sinusoidal applied electric potential is reasonably well-established. The mechanism at play is the same as that in the classical Mathieu equation (see ref.~\cite{singh2017levitation}) to describe dynamics under oscillatory forcing. A time average net ponderomotive force acting towards the center of the trap overcomes destabilizing forces such as gravity and charge repulsion (in case of two drops). It is then worthwhile to explore the robustness of the mechanism, by subjecting it to non-sinusoidal applied potentials. It is known that sinusoidal potentials of frequency $\omega$ can lead to responses in the deformation of the droplet in $\omega$ or $2 \omega$, which can be attributed to the quadrupole potential acting on the net charge or the square of the Maxwell stress tensor of the applied potential, respectively. From the requisite analysis of different waveforms, it is also known that non-sinusoidal potentials such as square and ramp consist of individual harmonics (odd, even or integer order multiplication of fundamental frequency) of the applied frequency. Thus it is interesting to see the impressions of these harmonics in the drop deformation pattern and establish the deformation response to arbitrary potentials. Undertaking a systematic investigation of this issue forms another motivation for the present study.

In the present manuscript, we have explored the surface oscillation characteristics in the presence of different forcing waveforms such as sine, square, and ramp. Unlike \citet{singh2018surface}, the surface oscillation characteristics are examined by the Fast Fourier transform (FFT) analysis of surface dynamics. The experimental observations are also compared with the asymptotic theory and boundary integral (BI) simulation in the potential flow limit. After a successful analysis of surface oscillations of a sub-Rayleigh charged (above Rayleigh charge the droplet surface becomes unstable even for small infinitesimal perturbations, see ref~\cite{duft03}) drop in the presence of non-sinusoidal applied potential we allowed the droplet to evaporate further, thereby, droplet attains Rayleigh critical charge and breaks via ejecting a thick jet. The breakup of a charged droplet is captured in the presence of sine, square and ramp applied waveforms.  

\section{Materials and method}
The experiments presented in this work involve the levitation of ethylene glycol (EG) charged droplets. A small amount of NaCl was added to increase the electrical conductivity ($\sigma$) of the droplet, and the conductivity was measured using a conductivity meter (Hanna instruments). The value of $\sigma$ was obtained in the range of $\sim$ 50-80 $\mu$S/$cm$. The viscosity ($\mu_i$) of EG was measured using Ostwald’s viscometer, and the corresponding measured value was 0.016 Ns/$m^2$. The surface tension ($\gamma$) of the EG droplet was measured using the pendant drop (DIGIDROP, model DS) method and confirmed with the spinning drop (dataphysics, SVT 20) apparatus. The value of $\gamma$ was obtained as 47 mN/m. The experiments were carried out at normal atmospheric conditions (1 atm pressure and $25^0$ C temperatures).
\begin{figure}[h!]
	\centering
	\includegraphics[width=\linewidth]{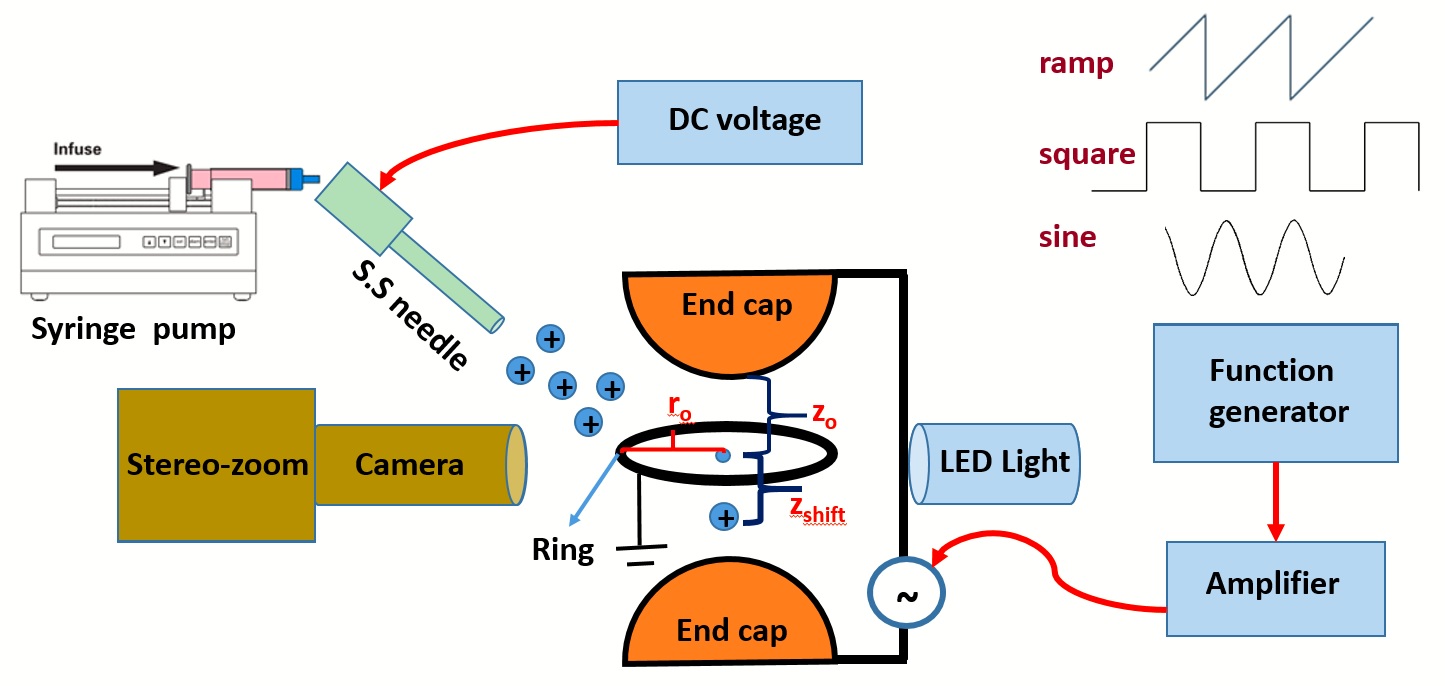}
	\caption{Detailed schematic of setup used for droplet charging and corresponding levitation of charged droplet.} \label{fig:setup}
\end{figure}

The experimental set-up consists of two end-cap electrodes (separation distance, $2z_0$= 12mm) and a ring electrode (diameter, $2r_0$= 12mm) to realise a quadrupole trap such that the electrical potential is given as $\phi=\Lambda(r^2-2z^2)$, where, $\Lambda$ (=$\frac{\phi_0}{r_0^2+2z_0^2}$) is the intensity of quadrupole field, $\phi_0$(=11 $kV_{pp}$) is the applied potential, $r$ and $z$ are the radial and axial directions respectively. A detailed schematic of the experimental setup used for droplet charging and levitation is shown in figure \ref{fig:setup}. A function generator (33220A Function /Arbitrary Waveform Generator, 20 MHz), used to apply the potential of the desired waveform, was connected to a high-voltage amplifier source (Trek, model 5/80, high-voltage power amplifier). The peak to peak AC potential applied in our experiments was 11$kV_{pp}$ with frequency varying from 100 Hz to 500 Hz. The voltage was kept highest and constant to ensure the high center of mass stability of a levitated charged droplet. In a typical experiment, charged EG droplets were generated by an electrospray (in dripping mode) setup, realized by applying high positive DC potential on the tip of a stainless steel needle. A charged droplet was further injected between the endcap and ring electrodes and was suspended using the quadrupolar AC electric field. The field was applied between the end caps and the ring electrodes of the trap where the end caps were connected to the live power supply, and the ring was kept grounded, as shown in figure \ref{fig:setup}.  
The levitated single droplet was observed using a high-speed CMOS camera (Phantom V 12, Vision Research, USA), which was connected with a stereo zoom microscope (SMZ1000, Nikon Instruments Inc.), see figure \ref{fig:setup}. The camera can record up to 180 thousand fps at 128$\times$128 resolution with 2s recording time and was kept inclined at $30^0$-$40^0$ for visualization of the phenomenon. Nikon halogen light (150 W) was used as a light source to illuminate the levitated droplet. The maximum frame rate which we can achieve with the camera at 128$\times$128 resolution is 130-150 hundred thousand fps, indicating a 130-150 kHz of frequency. The frequencies discussed in this work are of the order 100-500 Hz. Therefore, it suffices to say that the analysis is carried out for the frequencies lower than the frequency cut-off of the camera speed.  
 
\section{Results and discussion}
\subsection{Surface oscillations of a charged drop:} 
 In a typical levitation experiment, the gravitational force associated with the mass of the drop is balanced by imposing an additional DC bias voltage superimposed on the AC voltage (see ref.~\citet{duft03, duft02}). It should be noted that in the present experiments, only the AC field is applied for levitation of the charged droplet. However, if one levitated the droplet at the center of the trap by applying a DC bias voltage, the presence of an applied field will have a negligible effect on the droplet deformation. Since in most practical situations, such as electrospray, the droplet experiences an asymmetric electric field and non-zero gravity, the scenario is easily simulated in a more controlled way in quadrupole trap, if we do not have any additional DC voltage. It was observed that in the presence of gravity, the droplet gets levitated at an off-centered location in the downward z-direction (at a distance $z_{shift}$, as shown in figure \ref{fig:setup}) irrespective of the type of the applied waveform. An off-centered droplet experiences a local uniform field ($E$=$4\Lambda z_{shift}$) along with quadrupole field ($\Lambda$), and this local uniform field causes asymmetry of force on the droplet surface in the upward and downward directions due to finite body effect, i.e., $E \sim (z_{shift}\pm R)$, where, $R$ is the droplet radius. Thus, the presence of $z_{shift}$ modifies the stress distribution over the surface of the droplet and affects the oscillatory response.
\begin{figure}[t!]
	\centering
	\includegraphics[width=0.8\linewidth]{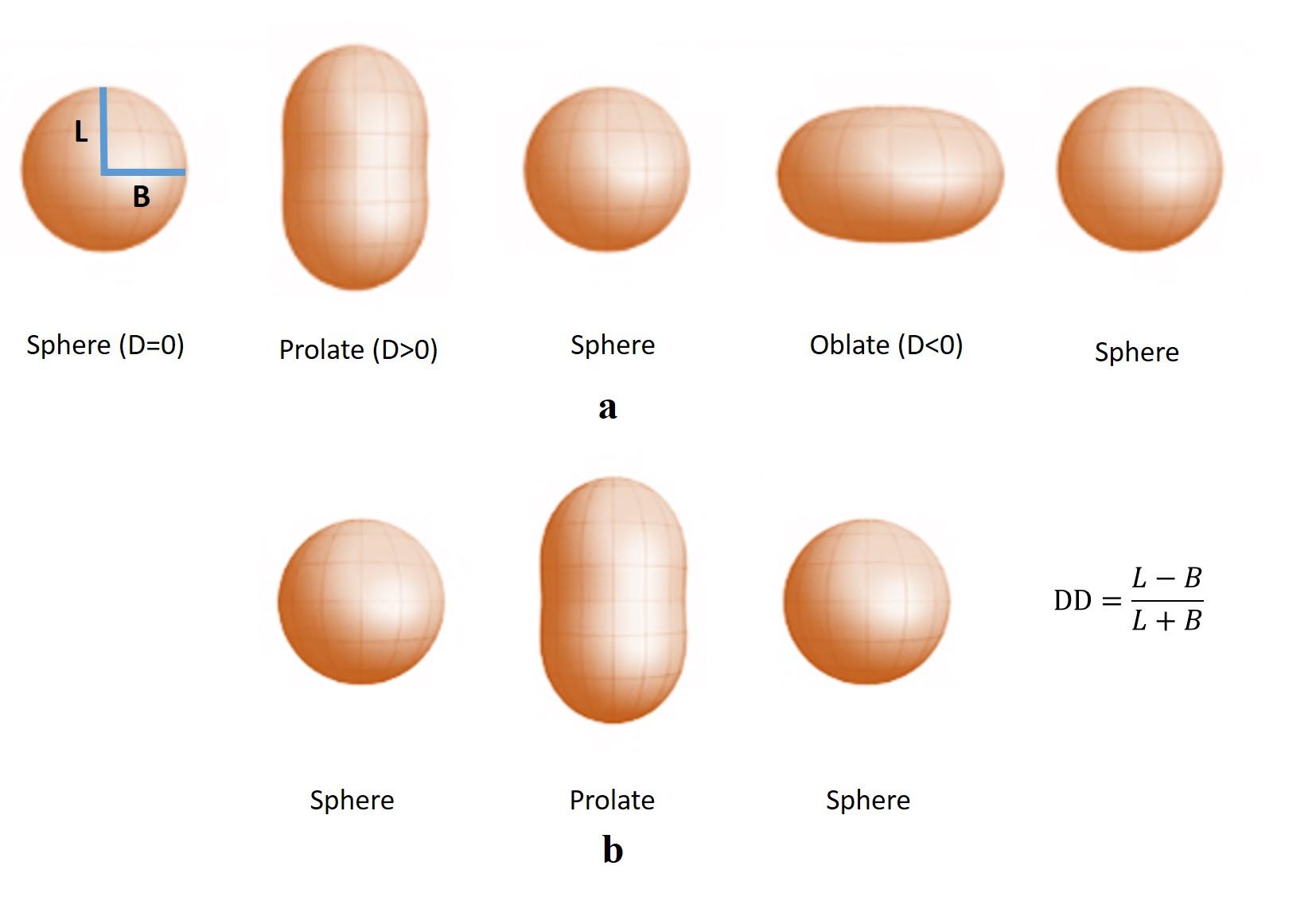}
	\caption{\textcolor{black}{Detailed schematic of SPSO and SPS modes of droplet oscillations.}} \label{fig:spso_sps}
\end{figure}
It was observed in some of our experiments that the droplet surface oscillates in the sphere-prolate-sphere-oblate (SPSO) mode. However, the magnitude of the prolate deformation is observed to be higher than that of the oblate deformation. This experimental observation is different from the experiments of \citet{duft02} wherein they observed symmetric SPSO mode and reported an equal magnitude of the prolate and oblate deformations in each oscillation cycle.A schematic of symmetric SPSO mode of droplet oscillations is shown in figure~\ref{fig:spso_sps}a. In some experiments, moreover, it was observed that the droplet exhibited a sphere-prolate-sphere (SPS) mode of oscillations, as shown in figure \ref{fig:spso_sps}b. Such oscillation pattern can be attributed to a complex interplay between the electrical parameters such as charge ($q$) on the droplet, intensity applied quadrupole field ($\Lambda$), local uniform field ($E$) as well as fluid properties such as viscosity, surface tension and density of the levitated droplet, and is discussed later. 

\subsection{Deformation under sinusoidal and other time periodic waveform} 

When a charged droplet is levitated using a quadrupole AC electric field with sine waveform as an applied signal, the droplet surface oscillates with the applied frequency. However, the surface oscillations can admit harmonic frequencies along with the fundamental applied frequency due to nonlinear interaction of several terms such as $q$, $\Lambda$ and $E$. In order to characterize experimental droplet surface oscillations, the high-speed video is processed using software ImageJ where the surface dynamics are obtained by tracking the variation in the outline of the droplet surface in each frame of the video. The accuracy of the analysis and noise in the data depends upon the quality of the images in each frame. It is reported by \citet{singh2018surface} that the voltage applied for levitation affects the amplitude of droplet surface oscillations. Thus, at a lower voltage, the amplitude of fundamental and harmonic frequencies becomes small, and thereby lower amplitude harmonics become undetectable. Hence, in the present experiments, the droplet is levitated at a constant and maximum AC voltage, i.e., 11$kV_{pp}$, and different types of waveforms are applied between the electrodes. Since droplet levitation is a result of dynamic stability and all the parameters are interdependent, we cannot apply a very high voltage or frequency. The experimental or theoretical analysis presented here is performed within the stability limit of the center of mass motion. The oscillation characteristics are examined by evaluating the rate of change of the Taylor deformation parameter ($DD$=$\frac{L-B}{L+B}$, where, $L$ is the major axis, and $B$ is the minor axis. The magnitude of $L$ and $B$ were obtained by tracking the boundary of the droplet using software ImageJ). The fundamental frequency and other harmonics are identified by performing the FFT analysis using the software OriginLab. The software automatically chooses the correct  sample interval of $\vartriangle$$t$ from the data. A triangular type of window is used to suppress frequency leakage, and the mean square amplitude method is used for power density normalization. The FFT analysis gives the amplitude of all kinds of frequencies present in the data. Since the high-speed video recording of $\sim$ 100-200 $\mu$m droplet diameter is performed at a low resolution, the blurriness in the image gives uncertainty in the exact drop size measurement. Additionally, the video was further processed using ImageJ software, where image greyscale and blur thresholding cause additional uncertainty in the measurement of various droplet dimensions such as major/minor axis and centroid calculations. The $DD$ data contains around 10-15\% average uncertainty. In the FFT analysis, the uncertainty in the DD vs time data can cause a slight reduction in the corresponding peak intensity while order and occurrence of the peaks are expected to remain the same. Hence in the present manuscript, the FFT analysis is found to be an accurate tool for characterization of experimental droplet surface oscillations and its comparison with theoretical/numerical results.
  
\begin{figure*}[tbp]
	\centering		
	\begin{subfigure}[b]{0.3\linewidth}
		\includegraphics[width=\linewidth]{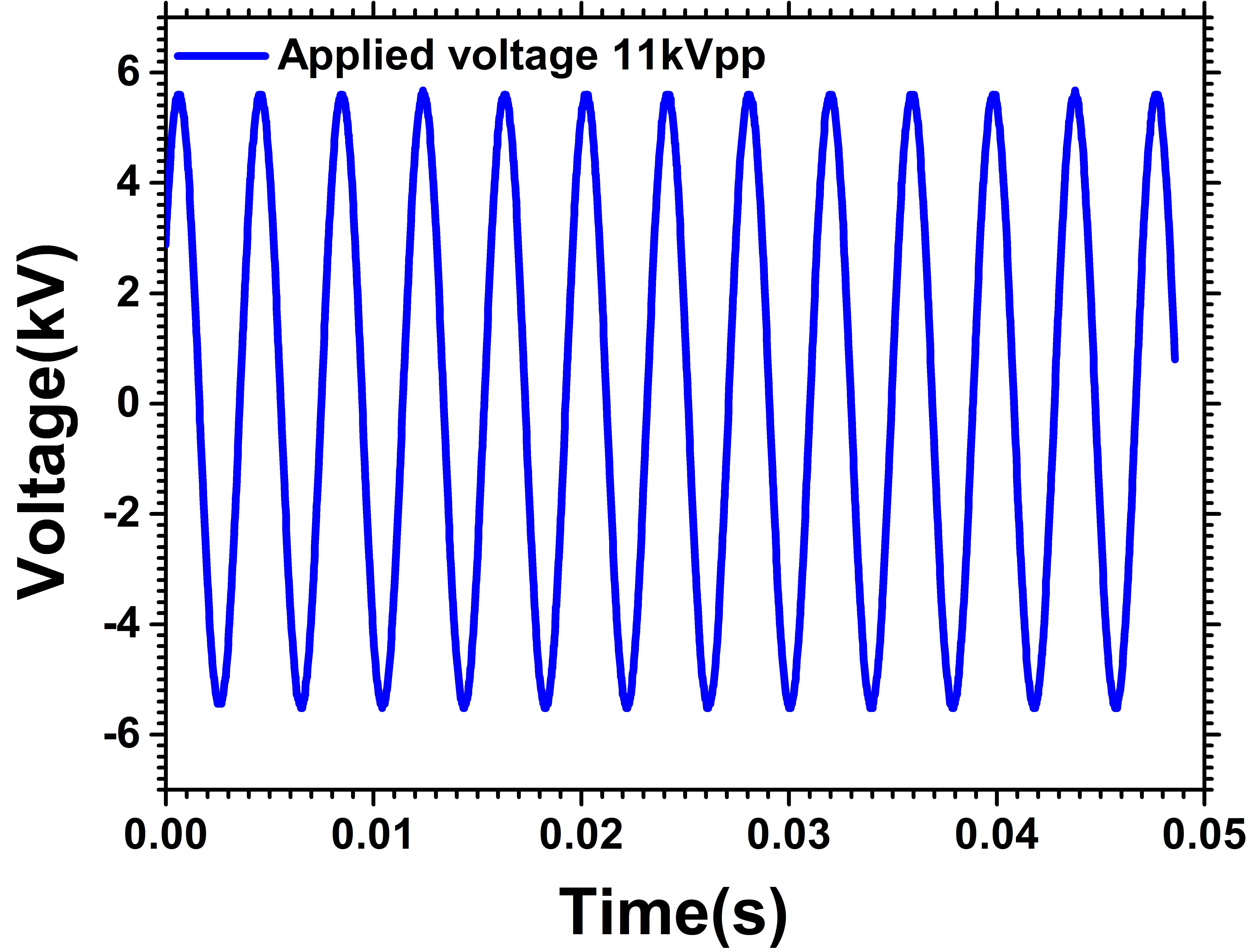}
		\caption{}
		\label{fig:applied_sin}
	\end{subfigure}
	\begin{subfigure}[b]{0.32\linewidth}
		\includegraphics[width=\linewidth]{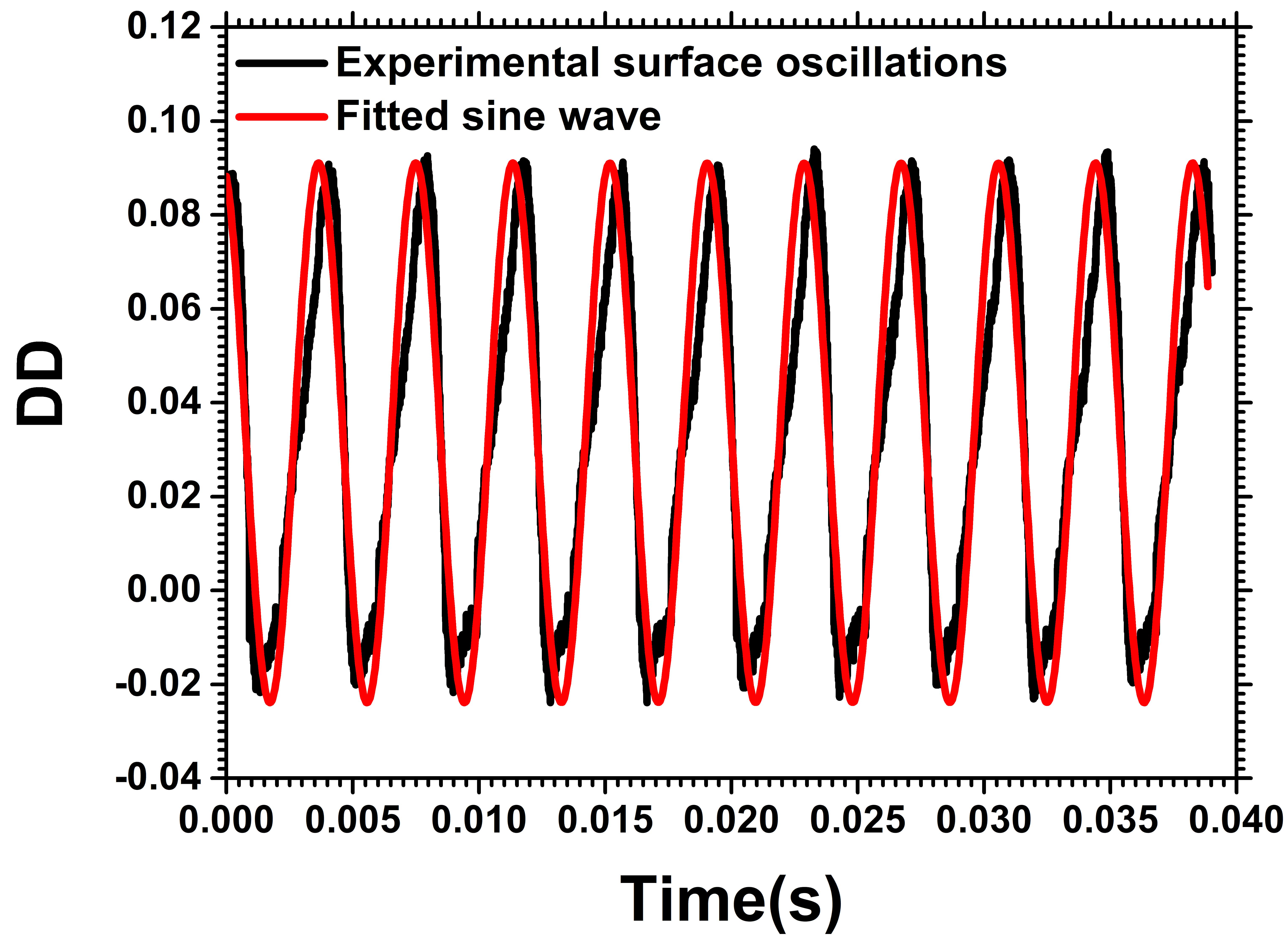}
		\caption{}
		\label{fig:sine_exp_fit}
	\end{subfigure}
	\begin{subfigure}[b]{0.32\linewidth}
		\centering
		\includegraphics[width=\linewidth]{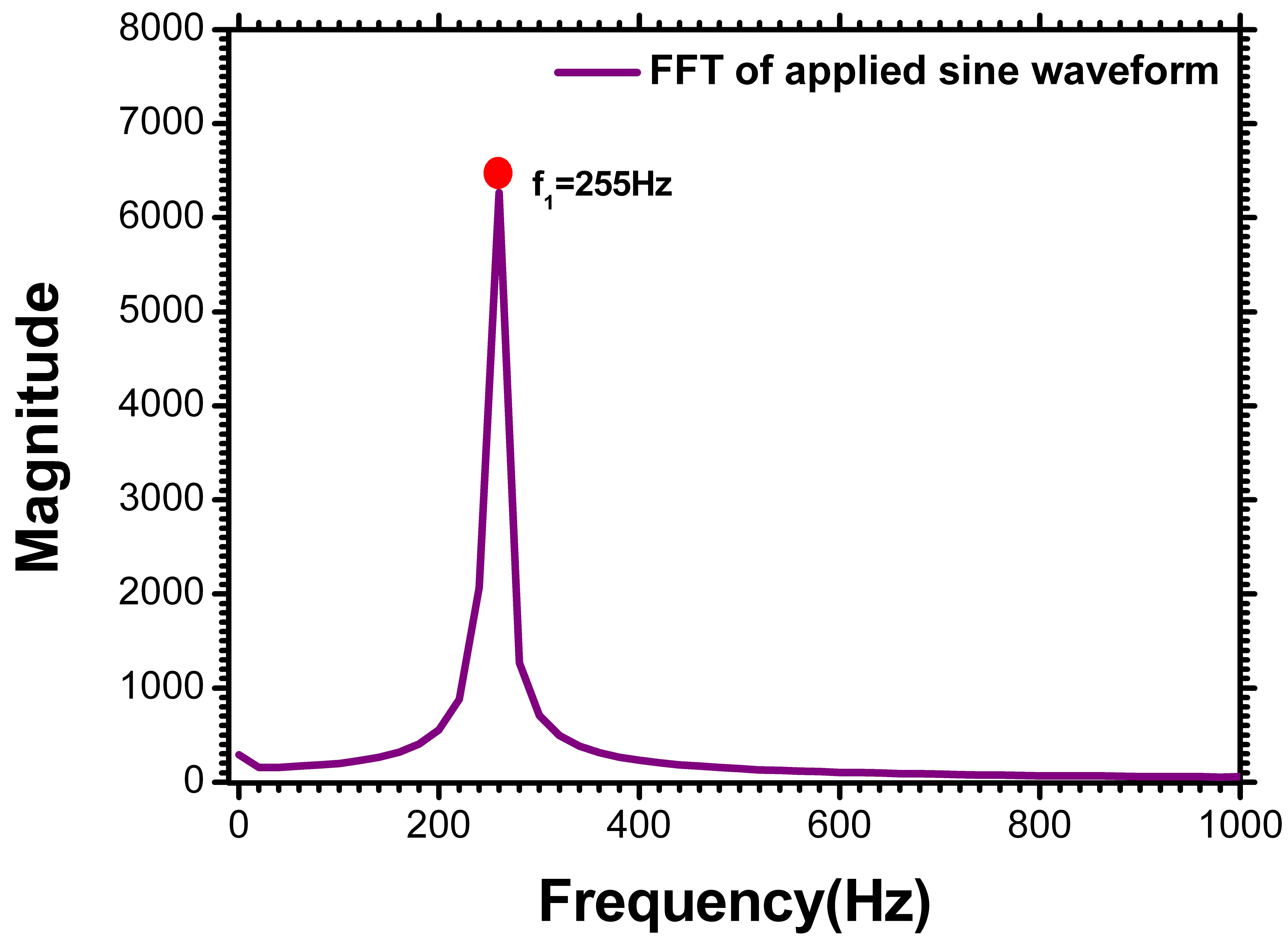}
		\caption{}
		\label{fig:Applied_sin_fft}
	\end{subfigure}
	\begin{subfigure}[b]{0.32\linewidth}
	\centering
	\includegraphics[width=\linewidth]{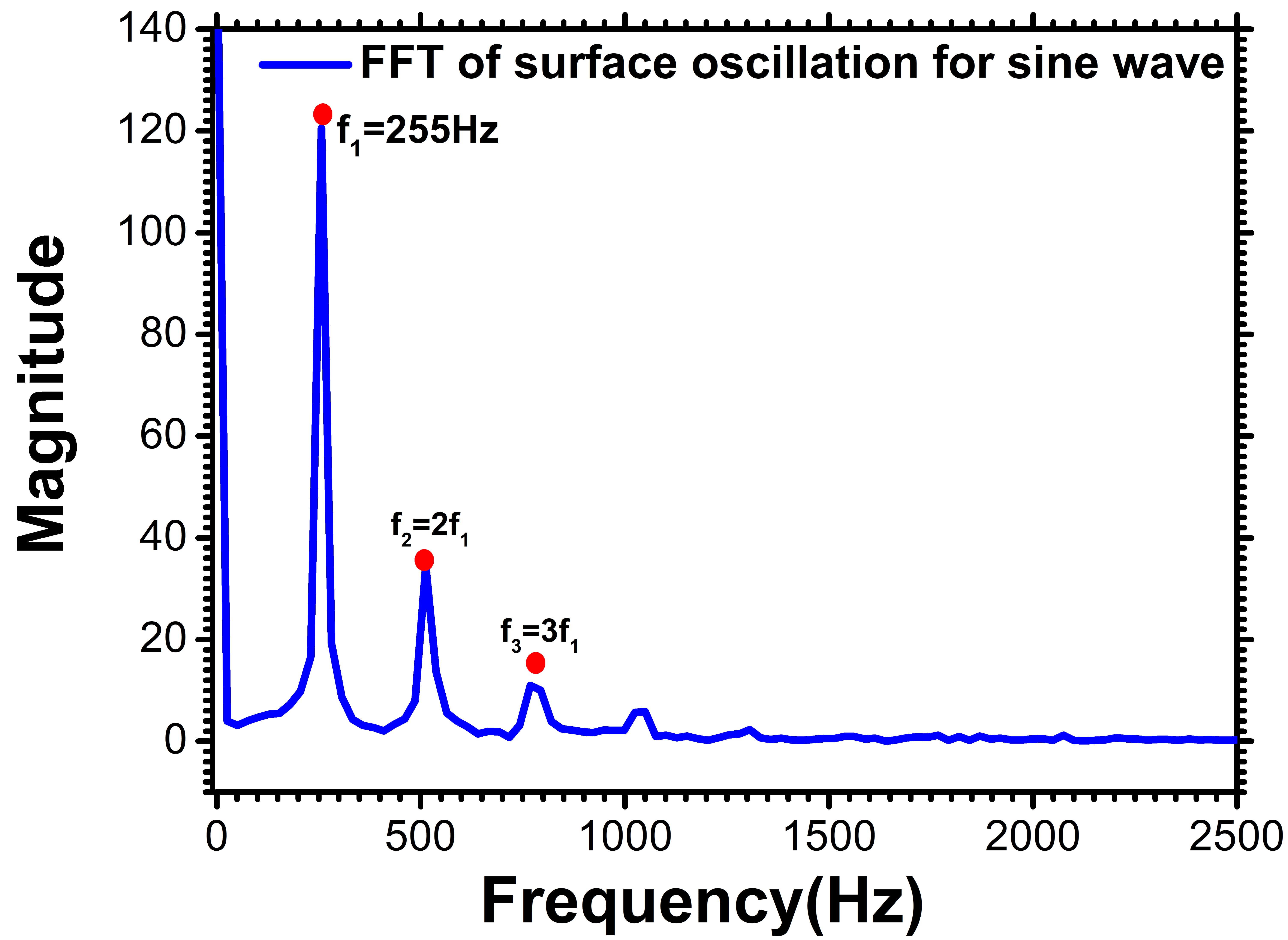}
	\caption{}
	\label{fig:exp_sin}
	\end{subfigure}
	\begin{subfigure}[b]{0.33\linewidth}
	\centering
	\includegraphics[width=\linewidth]{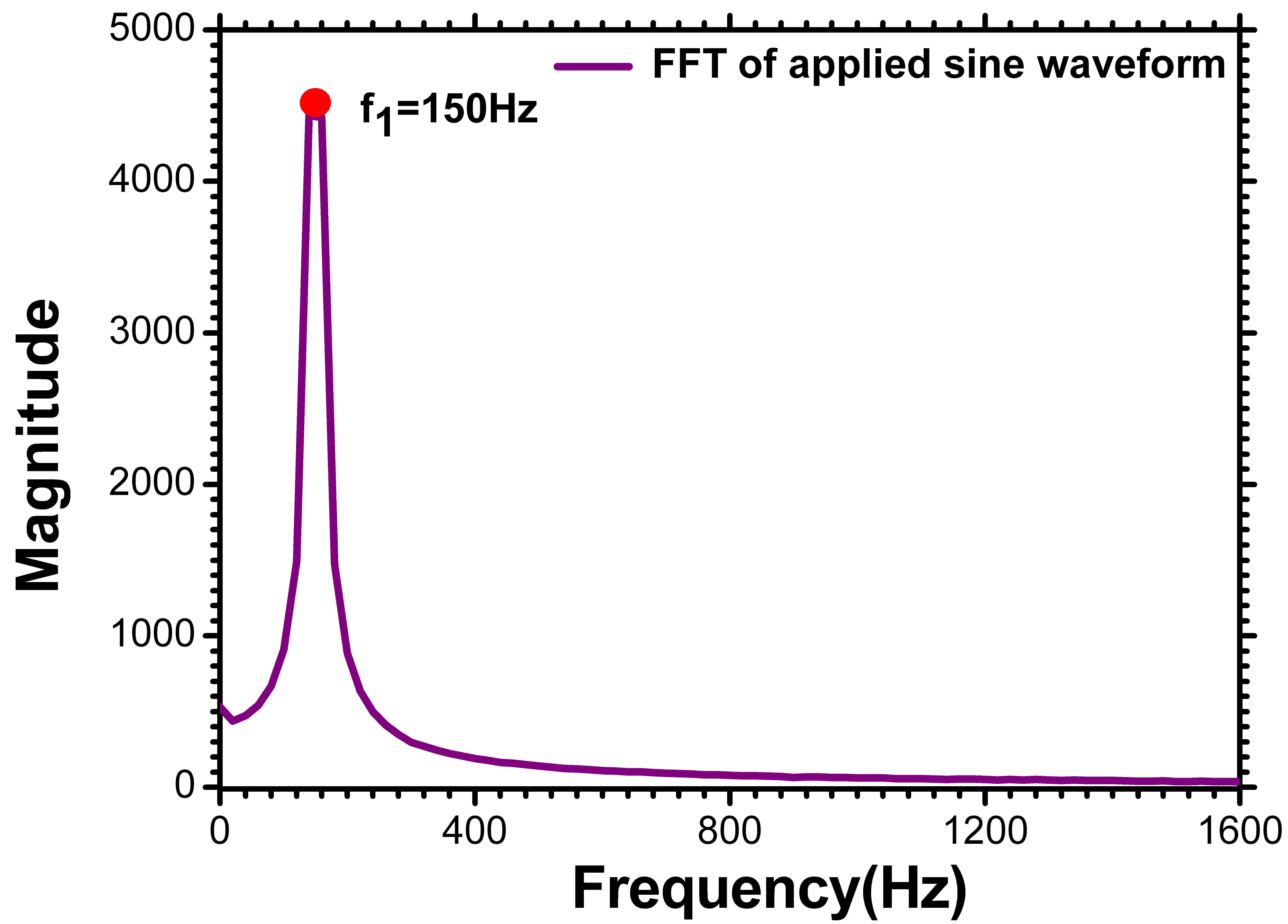}
	\caption{}
	\label{fig:applied_sin_new}
	\end{subfigure}
	\begin{subfigure}[b]{0.32\linewidth}
	\centering
	\includegraphics[width=\linewidth]{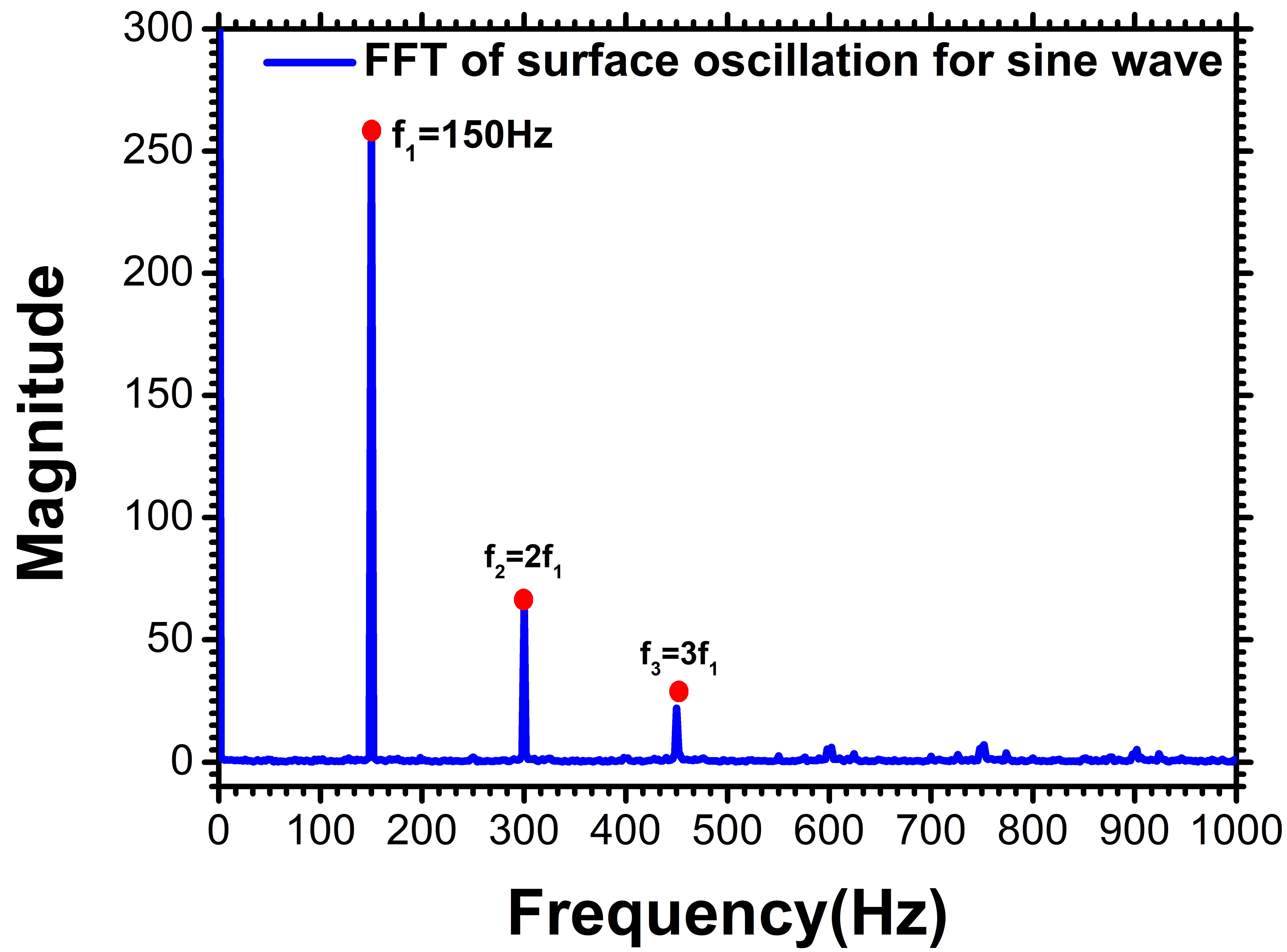}
	\caption{}
	\label{fig:exp_sin_new}
	\end{subfigure}
	\begin{subfigure}[b]{0.31\linewidth}
	\centering
	\includegraphics[width=\linewidth]{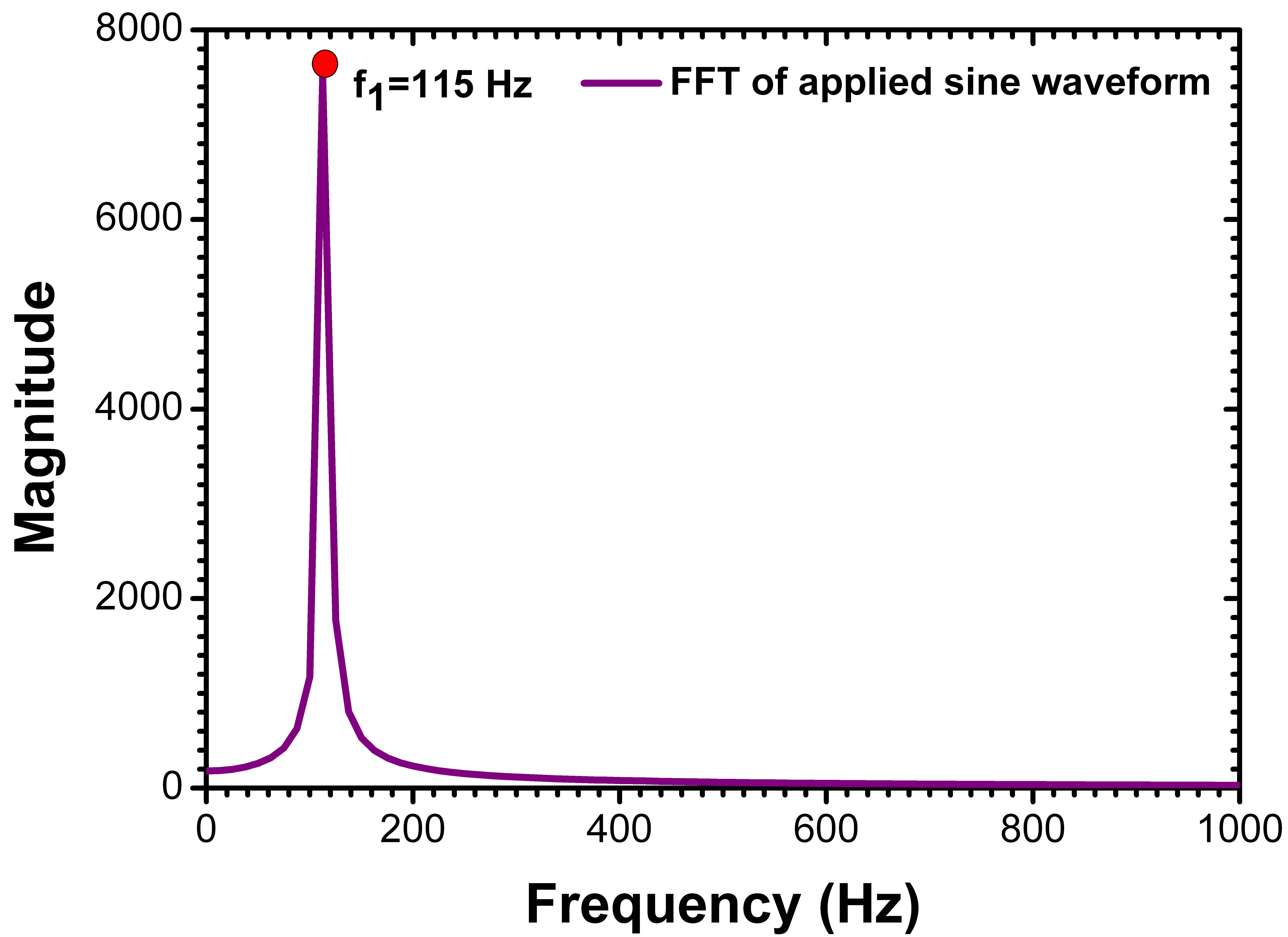}
	\caption{}
	\label{fig:applied_sin_new115}
	\end{subfigure}
	\begin{subfigure}[b]{0.3\linewidth}
	\centering
	\includegraphics[width=\linewidth]{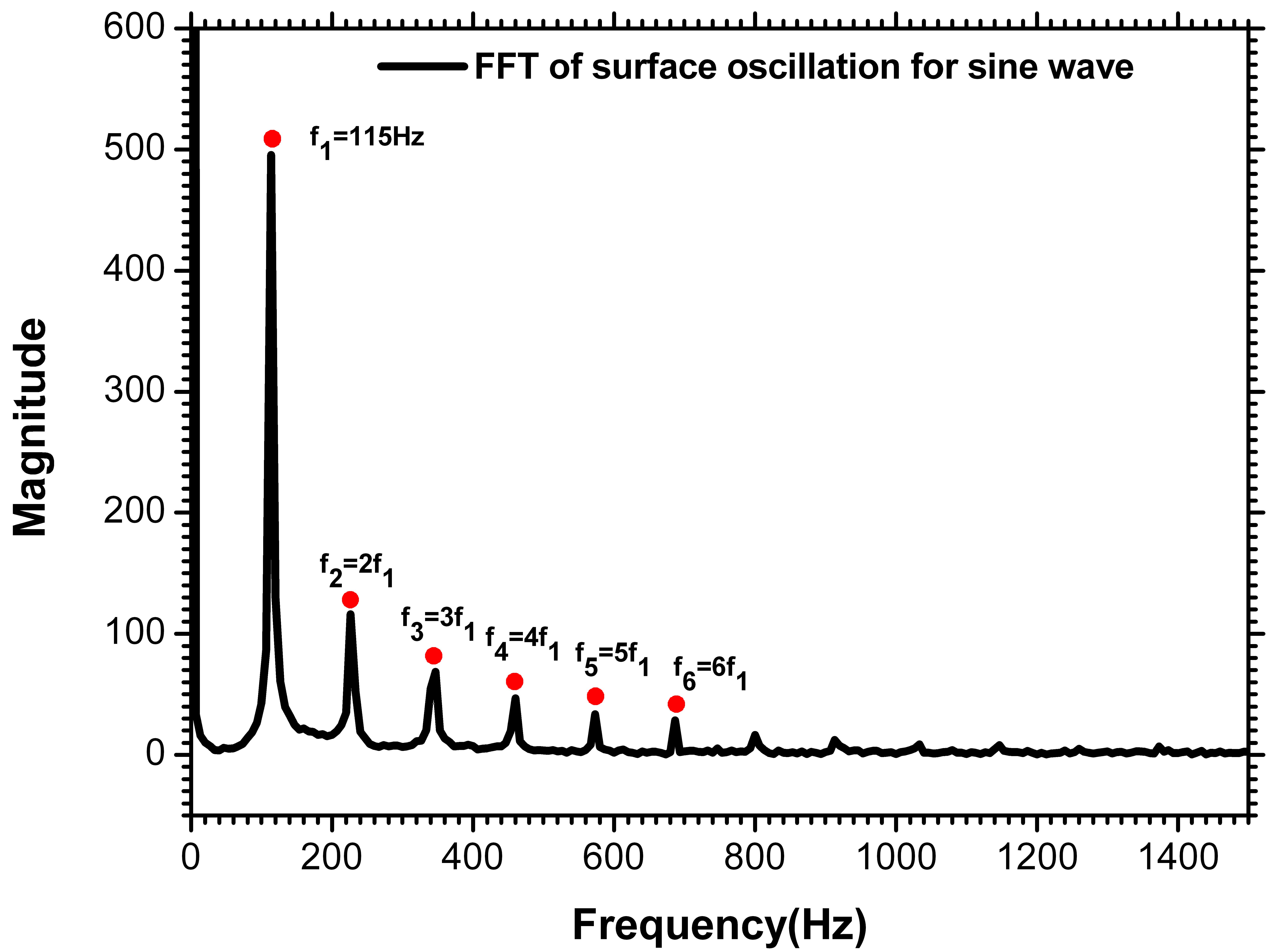}
	\caption{}
	\label{fig:exp_sin_new115}
	\end{subfigure}
	\caption{Droplet oscillation characteristics in the presence of a sine waveform; a) rate of change of applied voltage for sinusoidal waveform, b) surface oscillation of a droplet levitated in an ED balance, c) FFT analysis of sinusoidal applied signal used for levitating the droplet, d) FFT analysis of the DD vs time data e) FFT analysis of sinusoidal applied signal at 150Hz, f) FFT analysis of DD vs time data for 150 Hz applied frequency, (g) FFT analysis of sinusoidal applied signal at 115 Hz, h) FFT analysis of DD vs time data for 115 Hz applied frequency. Parameters; $D_d$=170$\pm$5 $\mu$m, $q$=1.0$\times$10$^{-11}$ C, fps=100k.}
	\end{figure*}

\citet{singh2018surface} reported that a droplet surface oscillates only with the fundamental frequency that is the applied frequency of the electric potential. However, it is apparent from the experimental data of DD vs time (figure 7 of \citet{singh2018surface}) as well as from figure \ref{fig:sine_exp_fit} of the present work that even with a perfect harmonic drive the droplet oscillation generates higher harmonics that are visible as a clear slanted deviation of the deformation signal from the fitted harmonic. A sine waveform of equal magnitude and frequency is fitted to the DD vs time data (see figure \ref{fig:sine_exp_fit}) using the software OriginLab for better depiction of the slanted nature of the curve. The detailed discussion of various factors responsible for slanted nature thereby harmonic oscillations is given the preceding section. Thus, to identify the harmonic frequencies, at first, a sinusoidal signal was applied between the endcap electrodes and a ring electrode to levitate a moderately charged droplet. Since at high value of applied frequency but at a fixed charge and applied voltage, the drop exhibits poor center of mass stability, thereby restricting the use of very high applied frequency close to the natural frequency ($O(2 kHz)$) of the system. Hence, the droplet can be stably levitated for a specific range of applied potential ($O(10 kV_{pp})$) and frequency ($O(200 Hz)$) that is determined by the levitation dynamics and critical frequency thereof. Thus, the first analysis is performed for a typical value of applied AC voltage of 11 k$V_{pp}$, and the corresponding applied frequency of 255 Hz. The data of the applied sinusoidal signal was acquired with the help of an oscilloscope (Tektronix) and is plotted in figure \ref{fig:applied_sin}.
The corresponding surface oscillations of a charged droplet is plotted in terms of DD vs time, as shown in figure \ref{fig:sine_exp_fit}. It can be observed from figure \ref{fig:sine_exp_fit} that the primary oscillation frequency of the droplet surface is the fundamental applied frequency, i.e., 225 Hz. However, an unmistakable noise at the peaks of the deformation cycle and slanting in the bottom curve can also be observed. This is an indication of the presence of higher harmonic frequencies. Hence, to confirm the presence of harmonic frequencies, an FFT analysis is performed for both applied waveform and the surface oscillation data. The FFT analysis of the applied signal is shown in figure \ref{fig:Applied_sin_fft}, and it can be observed that there exists a single peak at 255Hz which is the fundamental frequency of the applied signal. The FFT of surface oscillation on the other hand (shown in figure \ref{fig:exp_sin}) shows the existence of higher harmonics. Along with the fundamental frequency $f_1$=255Hz, the second harmonic frequency ($f_2$) at 2$f_1$ and the third harmonic frequency ($f_3$) at 3$f_1$ are clearly seen with diminishing magnitude. We hypothesized here that, while the fundamental ($f_1$) and second harmonics (2$f_1$) frequencies can be attributed to the effect of $\Lambda$ on $q$ and the $\Lambda^2$ term due to quadrupolar dependence of Maxwell stress on electric potential, the third harmonic could be due to the quadrupole field ($\sim$ $f_1$) acting on the charge induced on the deformed sphere ($\sim$ 2$f_1$) due to the quadrupole field itself. To re-confirm the presence of second and third harmonic frequencies, the droplet is levitated at a different frequency, i.e., at 150 Hz, and the FFT analysis is carried out for both applied signal and the surface oscillation data as shown in figures \ref{fig:applied_sin_new}, \ref{fig:exp_sin_new}. It can be observed from figures that while the FFT of the applied signal has only fundamental frequency (150 Hz) peak, as shown in figure \ref{fig:applied_sin_new}, the deformation admits harmonic frequencies, as shown in figure \ref{fig:exp_sin_new}. Additionally, the peaks in figure \ref{fig:exp_sin_new} seem to be much sharper than that of peaks observed in figure \ref{fig:exp_sin}. This is due to the better quality of the video thereby less noise in the DD vs time data. Also, it is believed that the non-linearities in the droplet oscillations due to large deformation can lead to the admittance of the $3^{rd}$ harmonic frequency. 
To confirm the results obtained at a lower applied frequency, i.e., 150 Hz, the droplet is levitated at a still lower frequency, i.e., 115 Hz and corresponding FFT analysis of both applied signal and DD vs time data are shown in figure \ref{fig:applied_sin_new115}, \ref{fig:exp_sin_new115}. At a very low frequency, for a fixed value of charge and applied voltage, the droplet
becomes unstable via spring oscillations of center of mass motion (see ref.~\citet{singh2017levitation}), whereby the freuency cannot be lowered any further. It can be observed that the FFT analysis of the applied signal, as shown in figure \ref{fig:applied_sin_new115}, shows the presence of a single peak, which corresponds to the applied frequency. On the other hand, the FFT analysis of droplet surface oscillation shows the presence of several harmonic frequencies along with fundamental frequency, i.e., 115 Hz. Its higher magnitude at 150 Hz as compared to that at 225 Hz is in agreement with the large deformation seen at the lower frequency.
\begin{figure*}[tbp]
	\centering		
	\begin{subfigure}[b]{0.38\linewidth}
		\includegraphics[width=\linewidth]{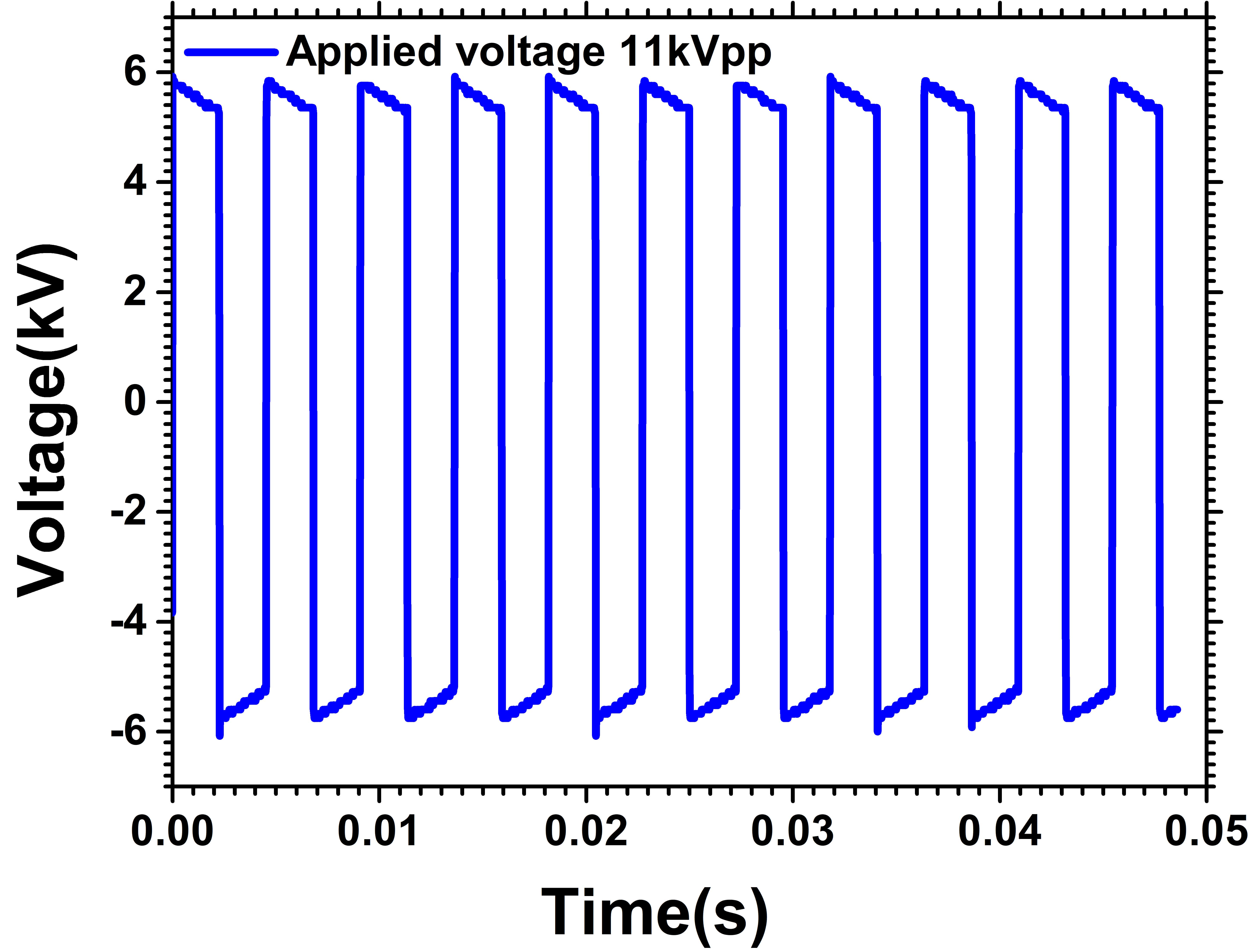}
		\caption{}
		\label{fig:applied_square}
	\end{subfigure}
	\begin{subfigure}[b]{0.38\linewidth}
		\includegraphics[width=\linewidth]{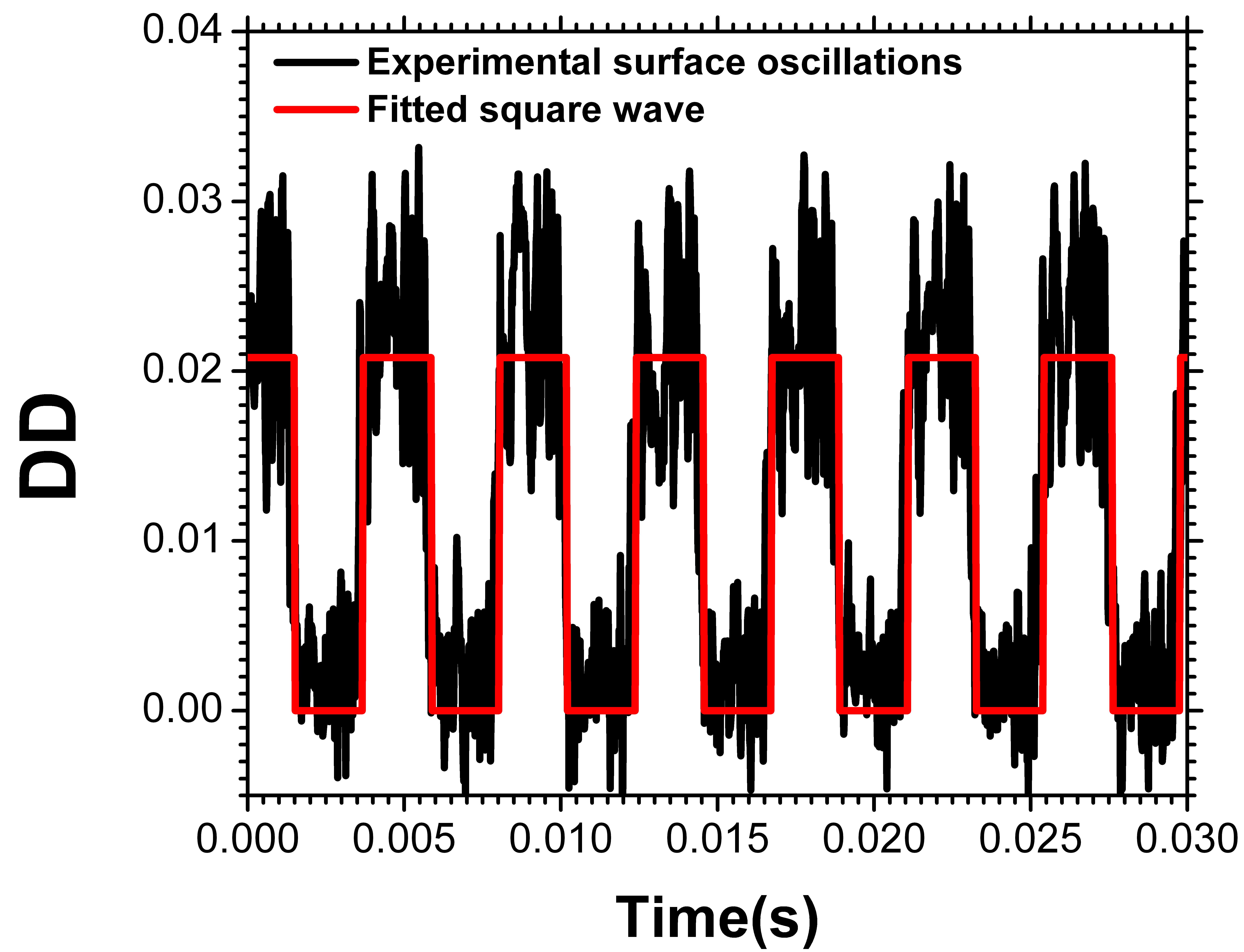}
		\caption{}
		\label{fig:square_exp}
	\end{subfigure}
	\begin{subfigure}[b]{0.42\linewidth}
		\centering
		\includegraphics[width=\linewidth]{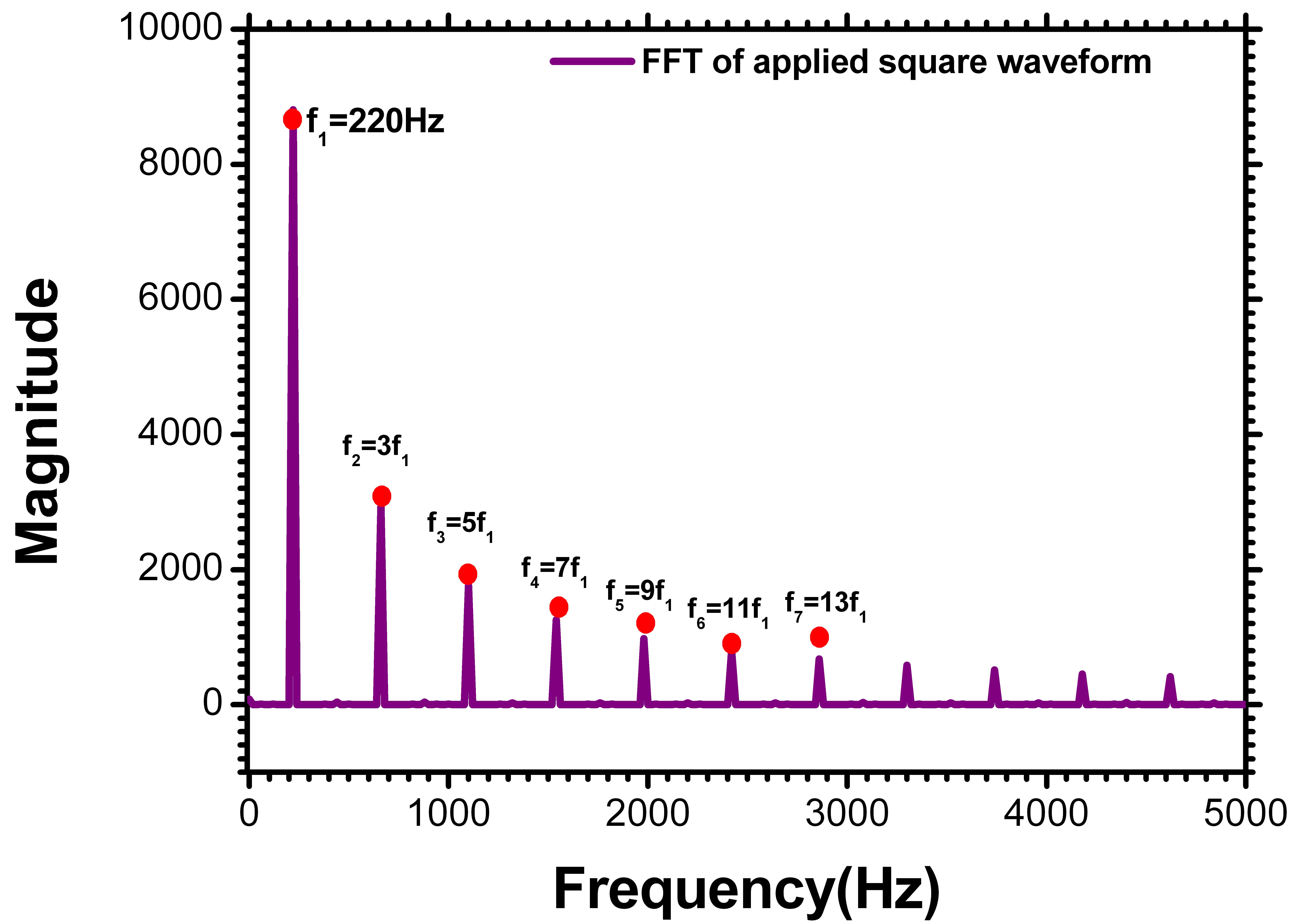}
		\caption{}
		\label{fig:applied_square_fft}
	\end{subfigure}
	\begin{subfigure}[b]{0.38\linewidth}
		\centering
		\includegraphics[width=\linewidth]{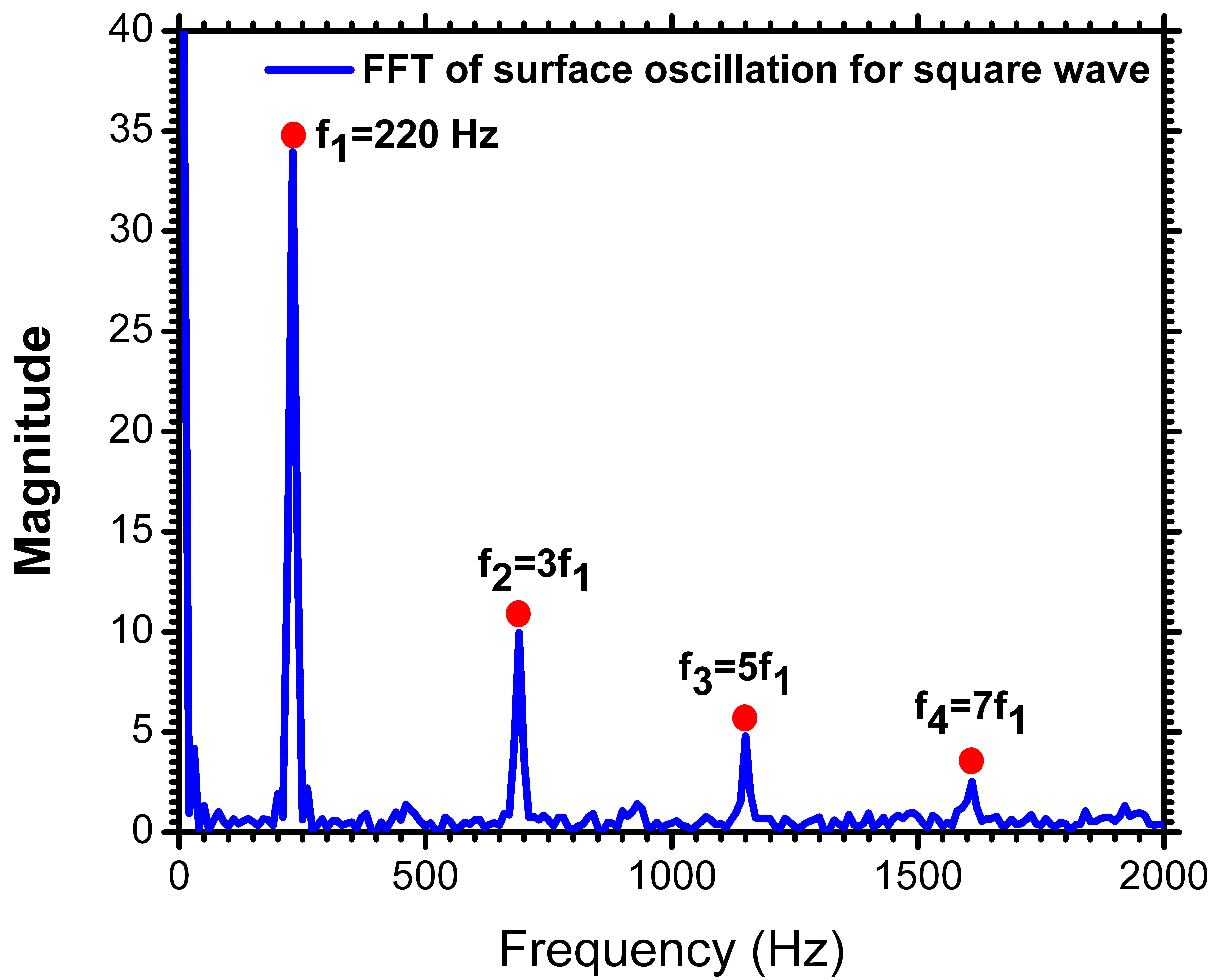}
		\caption{}
		\label{fig:exp_square}
	\end{subfigure}
	\caption{Droplet oscillation characteristics in the presence of a square waveform; a) rate of change of voltage for non-sinusoidal square wavefrom at 220Hz b) the surface oscillations of a droplet levitated in an ED balance in presence of square waveform c) FFT of square waveform applied signal for levitating the droplet d) FFT of degree of deformation vs time data for 220 Hz applied frequency. Parameters; $D_d$=110$\pm$5 $\mu$m, $q$=6.2$\times$10$^{-12}$ C, FPS=130k.}
\end{figure*}  

In the previous section, we have seen the levitation of a charged droplet in the presence of a sine waveform. 
 Therefore, two types of non-sinusoidal waveforms, namely square and ramp waveform, are applied for examining the controlled droplet surface oscillation characteristics. Firstly, to examine the case of the square waveform, a charged droplet is levitated at 11 k$V_{pp}$ applied potential and 220 Hz imposed frequency. The characteristics of the applied square wave signal, obtained from the oscilloscope, are shown in figure \ref{fig:applied_square}. The corresponding droplet surface oscillations, obtained from the image processing of high-speed video, are plotted in terms of DD vs time and shown in figure \ref{fig:square_exp}. The apparent noise observed at the peaks of the applied signal (figure \ref{fig:applied_square}) as well as in the DD vs time plot (figure \ref{fig:square_exp}) are a consequence of higher odd-harmonics typically associated with square waveforms \cite{weisstein2004fourier}. 
Additionally, the noise in the deformation data is a consequence of the noise in the applied signal as well as the noise associated with image processing of the high-speed video. For the preliminary confirmation of the surface oscillations in the presence of a square waveform, a square wave function of the same amplitude and frequency is fitted to the DD vs time data using the software OriginLab. It can be observed from figure \ref{fig:square_exp} that the surface oscillation characteristics of a droplet follow the applied waveform, and the droplet oscillates with fundamental applied frequency. Further confirmation of droplet oscillation behavior is done by performing the FFT analysis of the applied waveform and the surface oscillation data. The FFT analysis of the applied signal is shown in figure \ref{fig:applied_square_fft}, and it can be observed that the applied signal has several harmonic frequencies in odd multiples of the fundamental frequency ($f_1$=220Hz). This response is expected because of the characteristic behavior of the square waveform. The magnitude of the fundamental frequency is observed to be high as compared to the harmonic frequencies. The corresponding FFT analysis of surface oscillation behavior is shown in figure \ref{fig:exp_square}, and it can be observed that the FFT analysis of the surface oscillation behavior is similar to that of the applied signal. The order of occurrence of the harmonic frequencies is similar to the applied signal. 
Unlike the sinusoidal waveform, the droplet surface oscillates with applied frequency in the presence of square waveform, and the $2^{nd}$ harmonic is also significantly marked. The dominant $3^{rd}$ and $4^{th}$ harmonic in surface oscillations here is actually the response of the droplet to the presence of $3^{rd}$ and $4^{th}$ harmonic in the applied voltage.
\begin{figure*}[tbp]
	\centering		
	\begin{subfigure}[b]{0.38\linewidth}
		\includegraphics[width=\linewidth]{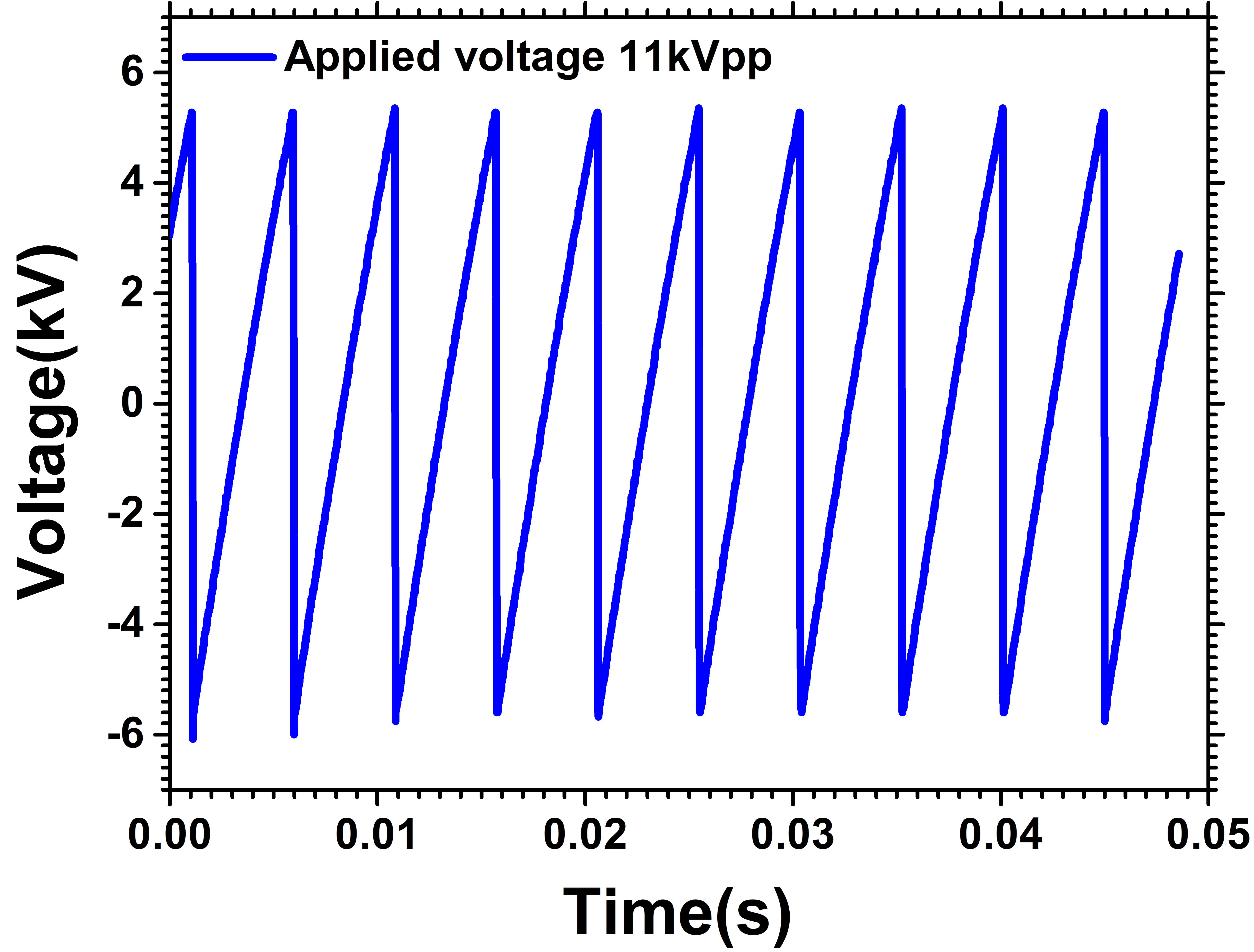}
		\caption{}
		\label{fig:apllied_ramp}
	\end{subfigure}
	\begin{subfigure}[b]{0.4\linewidth}
		\includegraphics[width=\linewidth]{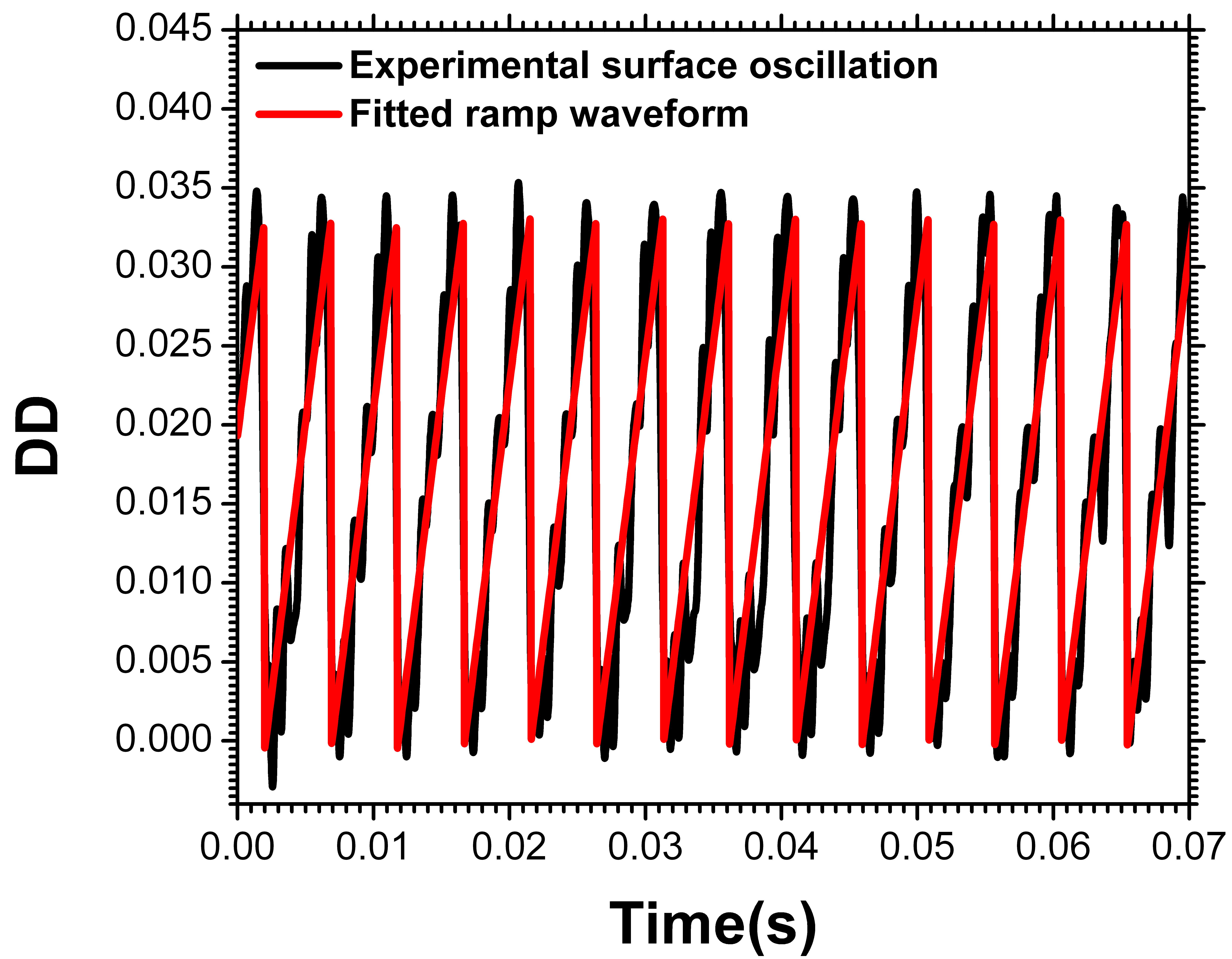}
		\caption{}
		\label{fig:triangular_exp}
	\end{subfigure}
	\begin{subfigure}[b]{0.4\linewidth}
	\includegraphics[width=\linewidth]{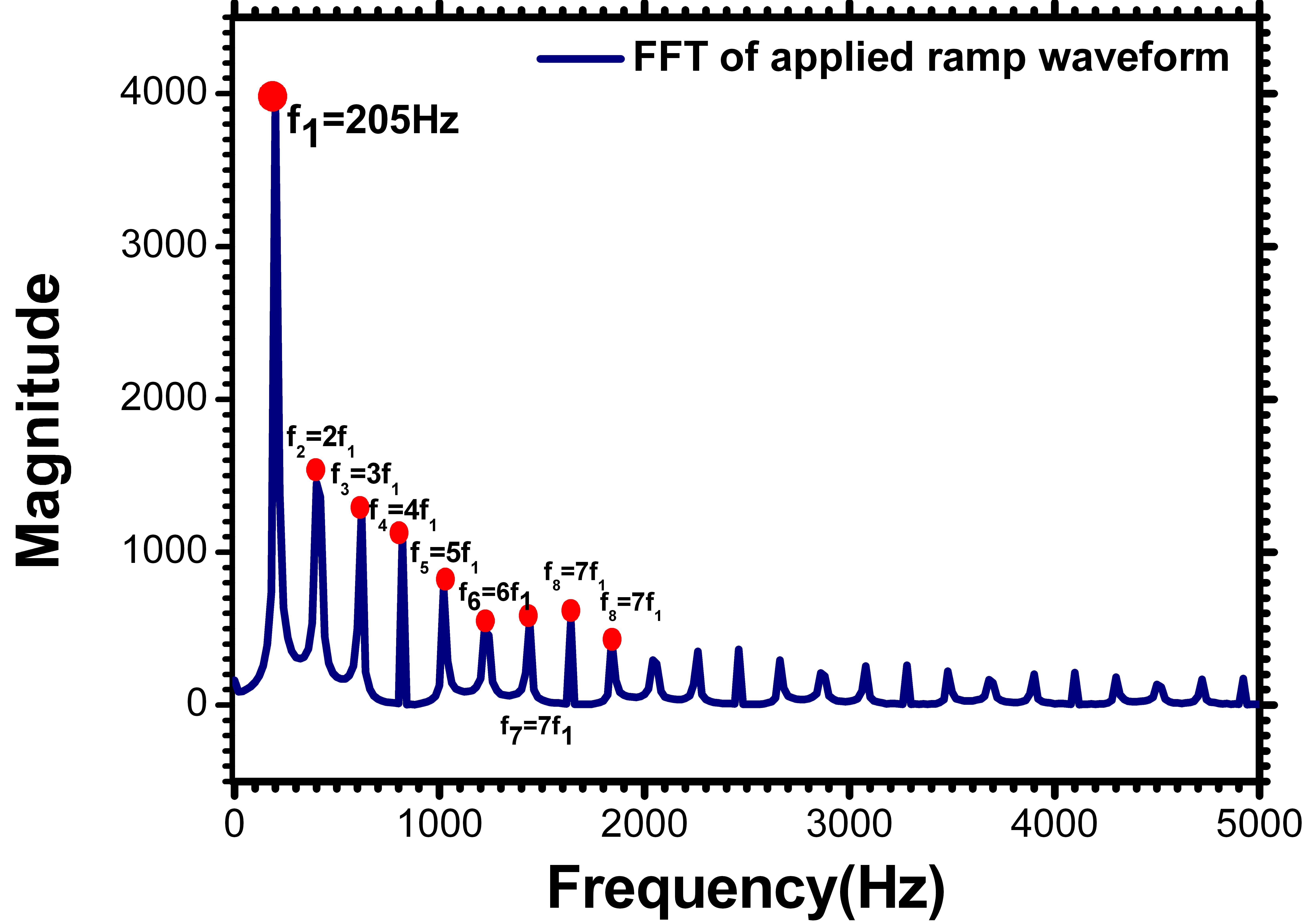}
	\caption{}
	\label{fig:applied_ramp_fft}
     \end{subfigure}
	\begin{subfigure}[b]{0.4\linewidth}
		\centering
		\includegraphics[width=\linewidth]{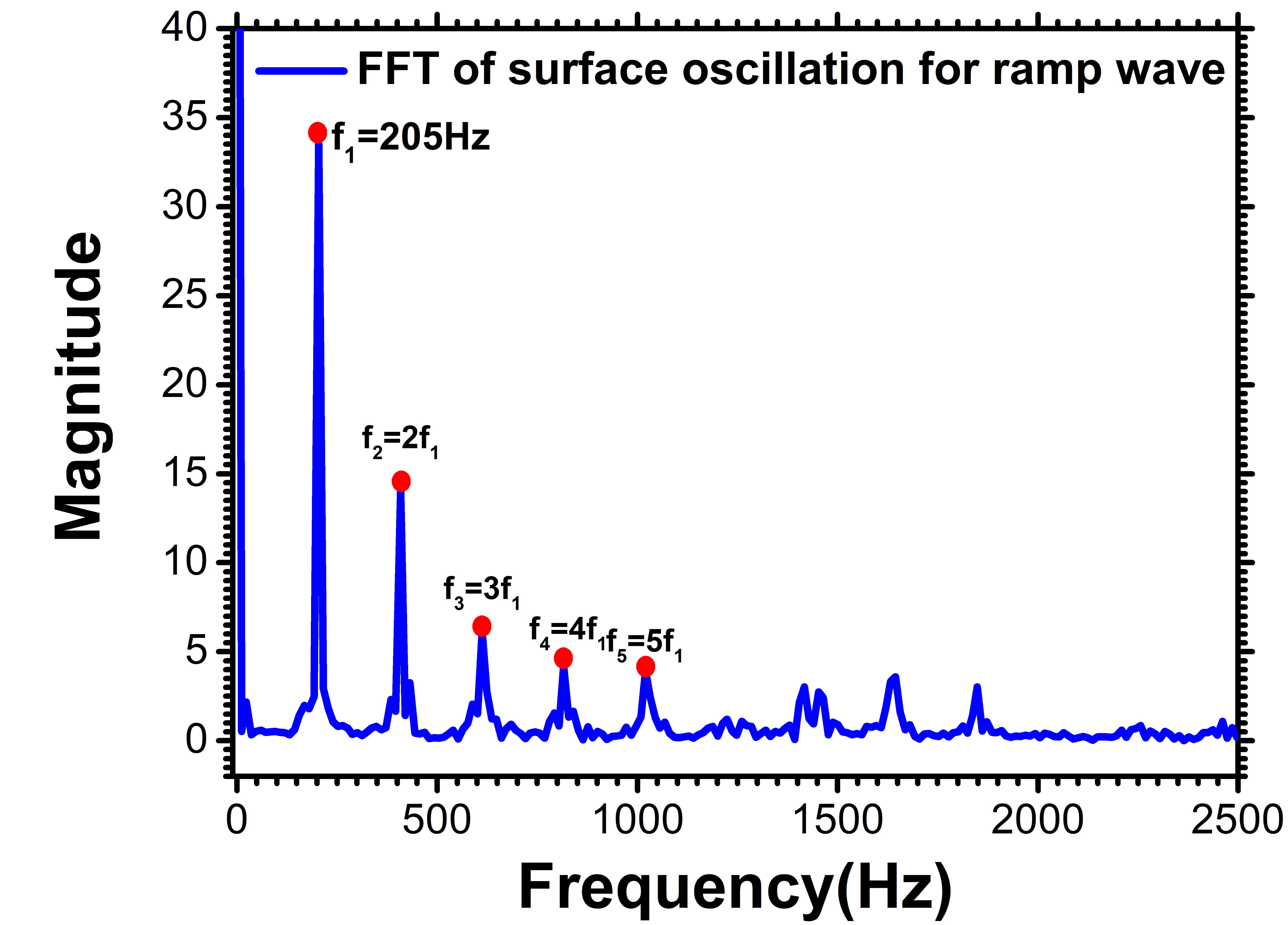}
		\caption{}
		\label{fig:exp_ramp}
	\end{subfigure}
	\caption{Droplet oscillation characteristics in the presence of a ramp waveform; a) rate of change of voltage for non-sinusoidal square waveform at 220Hz b) the surface oscillations of a droplet levitated in an ED balance in presence of square waveform c) FFT of square waveform applied signal for levitating the droplet d) FFT of degree of deformation vs time data for 220 Hz applied frequency. Parameters; $D_d$=120$\pm$5 $\mu$m, $q$=6.8$\times$10$^{-12}$ C, FPS=130k.} 
\end{figure*} 

Similar to the square waveform, another non-sinusoidal waveform, a ramp waveform, is applied to levitate a charged droplet in an ED balance. The voltage applied across the ring, and the end cap electrode is 11 k$V_{pp}$ at 205 Hz applied frequency. The rate of change of applied voltage with time for ramp waveform is shown in figure \ref{fig:apllied_ramp}. The surface oscillations in the form of DD vs time are shown in figure \ref{fig:triangular_exp}. Unlike the case of sine and square waveforms, it can be observed that the droplet oscillates in the SPS mode of oscillations. To confirm that the oscillations follow the applied waveform a ramp waveform of equal magnitude and frequency is fitted onto the DD vs time data using the software OriginLab and a good fit is observed, as shown in figure \ref{fig:triangular_exp}. Figure \ref{fig:applied_ramp_fft} shows the FFT analysis of the applied ramp waveform, and it can be observed that there exist integer harmonic frequencies in the applied signal. This is because the ramp waveform is theoretically an infinite series of sine integer harmonic frequencies \cite{weisstein1999fourier}. Although the magnitude of the fundamental frequency is higher than that of harmonics frequencies, all harmonic frequencies do show a significant contribution to the ramp signal. This is evident in figure \ref{fig:exp_ramp}, where the FFT of the deformation shows a significant presence of the fundamental and all higher harmonics. These results conclusively prove that an unmistakable signature of the applied waveform is seen on the deformation dynamics.   

\subsection{Theoretical model:} 
The surface oscillation dynamics of a charged droplet levitated in an ED balance is governed by both electrostatics and hydrodynamics. Recently, \citet{singh2018surface} conducted a linear stability analysis for surface oscillations of a levitated charged droplet in the presence of a sinusoidal applied potential. In the present work, we extended the theoretical analysis for non-sinusoidal waveforms, analyzed the non-linearity within the system, and compared the results with the experimental observations. The details of the model are omitted here, only boundary conditions and governing equations for the surface dynamics are shown for self-consistency of the manuscript. The value of $\sigma$ is high (\textgreater 30 $\mu$S/cm), and the surrounding medium (air) is a perfect dielectric medium while the droplet is considered as a perfect conductor drop. Thus, the non-dimensional equations and the boundary conditions governing the electrostatics of the droplet surface dynamics are:

\begin{equation}
{\bf{\nabla}}^2 \phi =0, \quad  0 \leq r \leq r_s(\theta,t)
\label{Epot}
\end{equation}
\begin{equation}
\phi\rightarrow \phi_\infty, \quad r\rightarrow \infty
\end{equation}
\begin{equation}
{\bf{n}}\cdot {\bf{\nabla}} \phi= -q(\theta,t), \quad r=r_s(\theta,t)
\label{norm_E}
\end{equation}
\begin{equation}
{\bf{t}}\cdot {\bf{\nabla}} \phi= 0, \quad r=r_s(\theta,t)
\label{eqn:tang_E}
\end{equation}
where, the droplet surface is defined with a small amplitude of perturbation around a spherical shape ($r_s(\theta,t)$) as,

\begin{equation}
r_s(\theta,t)=[1+\sum_{l}\delta \thinspace\alpha_{l}(t)\thinspace P_{l}\cos(\theta)]
\end{equation}
where, $\delta$ is a small perturbation parameter, $P_{l}$ is the $l^{th}$ Legendre mode and $\alpha$ is the amplitude of corresponding Legendre mode as well as measures the oscillation amplitude. The $l$=1 mode contributes to the center of mass translation of the drop and the other modes such as $l$=2, 3, 4 indicate the deformation of the drop shape from a sphere, e.g., $l$=2 gives dipolar shape deformation (dumbbell shape), $l$=3 gives asymmetric shape deformation (pear shape), $l$=4 gives quadrupolar shape deformations.    

\textcolor{black}{Typically, an ethylene glycol droplet of diameter ($D_d$) 100-250$\mu$m with viscosity ($\mu_i$)= 0.016 Ns/$m^2$, density ($\rho_i$)=1097 Kg/$m^3$ and surface tension ($\gamma$)=47 mN/m is levitated in an electrodynamic balance to study the oscillation characteristics of the drop surface.} The typical range of Ohnesorge number ($Oh$=$\frac{\mu_i}{\sqrt{\gamma \frac{D_d}{2}\rho_i }}$), which relates the viscous and surface tension forces, for our experimental parameters is about 0.15-0.35. \textcolor{black}{The value of $Oh$ number, i.e., $Oh$\textless1, indicates that for the given experimental conditions, inertial effects can be dominant while the contribution of viscous terms cannot be neglected. Thus, for the theoretical analysis of the oscillation behavior in the potential flow limit, the viscosity is included through the normal stress balance condition. The corresponding theoretical analysis is known as viscous potential flow theory} and is governed by the following non-dimensional equations:
 
\begin{equation}
\bnabla \cdot {\bf{v}}_{i,e}=0
\end{equation}
\begin{equation}
\frac{\beta}{Oh^2}
\Biggr{(}\frac{\partial v_{i,e}}{\partial t}+ v_{i,e}\cdot \bnabla v_{i,e}\Biggr{)}=-\bnabla p_{i,e}+\lambda_{i,e}\bnabla^2 v_{i,e}
\end{equation}
\textcolor{black}{Where, subscript $i$ and $e$ refer to drop and the outside medium (i.e., air) respectively, $\beta=\rho_e/\rho_i$ , where, $\rho_e$ and $\rho_i$ are the densities of the air and drop respectively, $\lambda=\mu_e/\mu_i$, where, $\mu_e$ and $\mu_i$ are the viscosity of the air and drop respectively.}
In the potential flow limit ($Oh\ll1$), the dynamics of the irrotational and incompressible fluid having velocity potential $\psi$ and the drop shape $r_s(\theta,t)$ is given by the following non dimensional equations, 
\begin{equation}
{\bf{\nabla}}^2 \psi_i =0, \quad  0 \leq r \leq r_s(\theta,t)
\label{velpot}
\end{equation}
\begin{equation}
{\bf{\nabla}}^2 \psi_e =0,  \quad r \geq r_s(\theta,t)
\end{equation}
and the pressure is given by the Bernoulli equation,
\begin{equation}
\triangle p+\Biggr{(}\frac{\partial \psi_i}{\partial t}+\frac{1}{2}\Biggr{[}{\frac{\partial \psi_i}{\partial r}}^2+{\frac{1}{r}\frac{\partial \psi_i}{\partial \theta}}^2\Biggr{]}\Biggr{)}-
\beta\Biggr{(}\frac{\partial \psi_e}{\partial t}+\frac{1}{2}\Biggr{[}{\frac{\partial \psi_e}{\partial r}}^2+{\frac{1}{r}\frac{\partial \psi_e}{\partial \theta}}^2\Biggr{]}\Biggr{)}=0
\label{bernoulli}
\end{equation}
The boundary conditions are normal velocity continuity, 
\begin{equation}
\frac{\partial \psi_i}{\partial r}=\frac{\partial\psi_e}{\partial r},
\label{vel_conti}
\end{equation}
and the normal stress balance, 
\begin{multline}
\triangle p-2( Oh \frac{\partial v_{r_i}}{\partial r}+\lambda_e Oh \frac{\partial v_{r_e}}{\partial r})\\=\kappa-\frac{1}{32} (8 X+12 \sqrt{\text{Ca}}\thinspace \zeta \cos (\theta )+5 \sqrt{\text{Ca}_Q}\thinspace \zeta (3 \cos (2 \theta )+1))^2,
\label{stress_bal}
\end{multline}
at the interface $r=r_s(\theta,t)$.
The equation \ref{stress_bal} is the non dimensional stress balance equation and $\kappa$ is the surface curvature and defined as:
	\begin{equation}
	\kappa=2+\delta\sum_{l=2,3,4}(l(l+1)-2)\alpha_l P_l(\cos \theta)
	\end{equation} 

The length is scaled by unperturbed droplet radius $R$(=$D_d/2$), velocity by $\sqrt{(\gamma/\rho R)}$, time by $\sqrt{(\rho R^3/\gamma)}$, $\zeta$ is a time-periodic function which depends on the type of applied waveform, e.g.,  $\zeta$=$\cos\omega t$ for the sinusoidal applied waveform. $X$ is the fissility, which is the ratio of the actual charge on the droplet to its Rayleigh charge ($q_R$=$\sqrt{(64 \pi^2 \epsilon_e R^3 \gamma)}$), where $\epsilon_e$ is the electrical permittivity of the surrounding fluid. $Ca_Q$=$(R^3 \epsilon_e \Lambda_{0}^{2}/\gamma)$ is the capillary number based on the strength of the applied quadrupole field and $Ca$=$(R\epsilon_e E^2/\gamma)$ is the electrical capillary number based on the uniform field, where $E$=$4\Lambda z_{shift}$. \textcolor{black}{Here the velocity and electric potentials are represented in terms of Legendre polynomial and the boundary conditions are solved to obtain following four differential equations.} 
\begin{equation}
\alpha_1''(t)+\frac{1}{2 \beta +1}\Bigr{(}12 \text{Oh}\thinspace \alpha_1'(t)-12 (\sqrt{\text{Ca}} \sqrt{\text{Ca}_Q}\thinspace \zeta^2+X \sqrt{\text{Ca}}\thinspace \zeta)\Bigr{)}=0 \label{eqn:pot_diff1}
\end{equation}
\begin{equation}
\alpha_2''(t)+\frac{6}{3 \beta +2} \Big{(}2 (\lambda +4) \text{Oh} \thinspace \alpha_2'(t)+4 \alpha_2(t)\\-3 \text{Ca}\thinspace \zeta^2-10 X \sqrt{\text{Ca}_Q}\thinspace \zeta-\frac{25 \text{Ca}_Q\thinspace \zeta^2}{7}\big{)}=0 \label{eqn:pot_diff2}
\end{equation}
\begin{equation}
\alpha_3''(t)+\frac{1}{4 \beta +3}\Bigr{(}24 (2 \lambda +5) \text{Oh}\thinspace \alpha_3'(t)\\+120 \alpha_3(t)-108 \sqrt{\text{Ca}} \sqrt{\text{Ca}_Q}\thinspace \zeta^2\Bigr{)}=0 \label{eqn:pot_diff3}
\end{equation}
\begin{equation}
\alpha_4''(t)+\frac{1}{5 \beta +4}\Bigr{(}840 (\lambda +2) \text{Oh} \thinspace\alpha_4'(t)\\+2520 \alpha_4(t)-900 \text{Ca}_Q\thinspace \zeta^2\Bigr{)}=0 \label{eqn:pot_diff4}
\end{equation}
Equation \ref{eqn:pot_diff1} is based on the $P_1$ Legendre mode and primarily contributes in the translational motion of the droplet. The equation \ref{eqn:pot_diff2} is based on the $P_2$ Legendre mode and similarly, equation \ref{eqn:pot_diff3} and \ref{eqn:pot_diff4} are obtained as coefficients of the  $P_3$ and $P_4$ Legendre modes. This infers that equations \ref{eqn:pot_diff2}, \ref{eqn:pot_diff3}, \ref{eqn:pot_diff4} govern the surface oscillation dynamics of the drop. From equation \ref{eqn:pot_diff2}, it can be observed that for a highly charged drop, the value of X$\sqrt{Ca_Q}$ term dominates over other terms, and the droplet oscillates with the fundamental frequency. On the other hand, for mildly or uncharged droplets levitated at the center of the trap, the term $Ca_Q\zeta^2$ dominates over other terms, and the droplet oscillates with twice the frequency of fundamental applied frequency. Hence, a lower oblate deformation as compared to the prolate deformation during the negative and positive cycle of endcap potential in the SPSO mode of surface oscillation of a positively charged drop levitated in a quadrupole field depends upon the relative magnitude of $Ca\zeta^2$, X$\sqrt{Ca_Q}\zeta$, $Ca_Q\zeta^2$. The equation \ref{eqn:pot_diff3} shows that at a very high value of $z_{shift}$, the droplet shape can become highly asymmetric. The equation \ref{eqn:pot_diff4} shows that at high value of the applied field, the droplet oscillates with twice the imposed frequency. The relative magnitudes of the shape coefficients such as $\alpha_2$, $\alpha_3$, $\alpha_4$ are obtained by solving equations \ref{eqn:pot_diff2}, \ref{eqn:pot_diff3} and \ref{eqn:pot_diff4} and these equations, finally, governs the characteristics of the droplet surface oscillations. For our experimental parameters, the magnitude of $\alpha_2$ found to be dominating over $\alpha_3$ and $\alpha_4$. Thus, the dynamics of a charged droplet oscillation in the presence of sinusoidal and non-sinusoidal waveform is explained using equation \ref{eqn:pot_diff2}. 

In the presence of gravitational force, it is observed that all the four differential equations (equations \ref{eqn:pot_diff1}-\ref{eqn:pot_diff4}) get coupled at higher order. Thus, the surface dynamics of the droplet which is governed by equations \ref{eqn:pot_diff2}, \ref{eqn:pot_diff3} and \ref{eqn:pot_diff4} is significantly affected by the equation \ref{eqn:pot_diff1} which gives the COM motion of the droplet. This indicates that the shape deformation dynamics in the presence of gravity is the result of nonlinear interaction of the gravitational force with the electro-capillary forces, which cannot be explained using linear order theory. Thus to capture these nonlinear effects, the problem is also solved using the boundary integral method in the potential flow limit.

\subsection{Boundary integral equations}

As the flow is assumed to be irrotational inside the drop, the velocity is given by the gradient of the velocity potential $\tilde{\v}=\tilde{\bnabla}\tilde{\psi}$ such that velocity potential follows the Laplace equation and the pressure fields are related to flow fields through unsteady Bernoulli equation as follow:
\begin{align}
\frac{\del \tilde{\psi}}{\del \tilde{t}}+\frac{1}{2}(\tilde{\bnabla} \psi \cdot \tilde{\bnabla} \tilde{\psi})+\frac{\tilde{p}}{\rho_i}&=0 \label{eq:pot_bernoulli}
\end{align}
The pressure in the equation \ref{eq:pot_bernoulli} can be obtained from the normal stress jump across the surface of the drop as shown below,
\begin{equation}
\tilde{p}=\tilde{p}^{ext}+\gamma \tilde{\bnabla}_s \cdot \n 
\end{equation}
where $\n$ is the outward unit normal and $\bnabla_s \cdot \n$ gives the curvature of the drop denoted by $\kappa$. For electrified drops under the action of gravitational force, 
\begin{equation}
\tilde{p}^{ext}=\rho_i \thinspace g \thinspace \tilde{z}+\frac{1}{2} \epsilon_e \tilde{E_n}^2
\label{eqn:pressure}
\end{equation}
where, first term is the force acting on the drop due to gravity and the second term is the electric stress acting on the surface of the drop.

Note that the dimensional quantities are indicated by tilde and non-dimensional quantities are without tilde. Here the time is non-dimensionalized by the characteristic inertial timescale, $\sqrt{\rho_i R^3/\gamma}$, the pressure by $\gamma/R$ and the total surface charge is non-dimensionalized by $\sqrt{\gamma R^3 \epsilon_e}$ such that the non-dimensional Rayleigh charge is $8\pi$. \textcolor{black}{Other quantities are scaled similar to that of linear order theory which is discussed in the previous section.} Thus, the non-dimensional governing equations for the flow field inside the drop are given by,
\begin{align}
\bnabla^2\psi&=0,\\
\frac{\partial \psi}{\partial t}+\frac{1}{2}\v \cdot \v&=-\kappa-Bo \thinspace z+\frac{1}{2}E_n^2
\end{align} 
where, $Bo=\rho_i g R^2/\gamma$ is the gravitational bond number.
\textcolor{black}{The integral equation for electric potential are well documented and can be found in \cite{gawande2017numerical}.} Here, only the hydrodynamic integral equations in the potential flow limit are discussed. According to classical potential flow theory the velocity potential can be expressed as a surface distribution of dipole density per unit area $\mu$. This is also known as double layer potential representation and is given as,
\begin{equation}
\psi({\xo})=\int  \mu(\x) \n \cdot \bnabla \G dA 
\end{equation}
The potential is discontinuous across the surface of the drop.When $\x$ approaches the surface from inside the drop,
\begin{equation}
\psi_1({\xo})=\frac{1}{2}\mu ({\xo})+\int_{PV} \mu (\x) \n \cdot \bnabla \G dA,
\label{eqn:psi1}
\end{equation}
and when $\x$ approaces the surface from outside the drop,
\begin{equation}
\psi_2({\xo})=-\frac{1}{2}\mu ({\xo})+\int_{PV} \mu (\x) \n \cdot \bnabla \G dA,
\label{eqn:psi2}
\end{equation}
such that the dipole density $\mu=\psi_1-\psi_2$. The normal component of the velocity is continuous across the drop surface while tangential component of the velocity is discontinuous. Thus the surface distribution of dipoles is equivalent to a vortex sheet and the circulation density $\gamma_s$ can be written as,
\begin{equation}
\gamma_s=-\n \times \bnabla_s \mu
\end{equation}

As introduced by Lundgren and Monsour (1988), a vector potential $\mathcal{A}$ is related to the velocity by,
\begin{equation}
\v=\bnabla\times \mathcal{A}
\end{equation}
such that,
\begin{align}
\mathcal{A}&=-\int\gamma_s \G dA\\
\mathcal{A}&=-\int_{PV} \mu(\x) \n \times \bnabla_s \G dA
\label{eqn:vectorPot}
\end{align}
Eq. \ref{eqn:vectorPot} is used to calculate the vector potential $\mathcal{A}$ when the point $\x$ is precisely on the drop surface. Thus the normal component of the velocity is then calculated using,
\begin{equation}
\v \cdot \n=(\n \times \bnabla) \cdot \mathcal{A}
\end{equation} 
For axisymmetric problem ($\n \times \bnabla$) operator has surface values only, thus the normal component of the velocity can be obtained once the vector potential $\mathcal{A}$ is computed. For initially known scalar potential on the drop surface the tangential velocity is obtained by calculating the surface gradient of $\psi$. Thus the tangential and normal component of velocity are given by,
\begin{align}
v_n&=\frac{1}{r}\frac{\del (r \mathcal{A}_\theta)}{\del s}\\
v_t&=\frac{\del \psi}{\del s}
\end{align}
Since the velocity of the drop surface is computed, the surface can be evolved in time using kinematic condition,
\begin{equation}
\frac{d \x}{d t}=\v
\end{equation}
and the surface values of $\psi$ at next time iteration are obtained by using unsteady Bernoulli equation in terms of the material derivative.
\begin{equation}
\frac{D \psi}{Dt}=\frac{1}{2} \v \cdot \v-\kappa-Bo \thinspace z+\frac{1}{2}E_n^2
\label{eqn:bernoulli}
\end{equation} 

The other details of the integral method, regularization of the kernel in equations \ref{eqn:psi1}, \ref{eqn:psi2} and \ref{eqn:vectorPot} and their numerical implementation can be found in the supplementary file. \textcolor{black}{It should be noted here that in BIM, viscous damping can be included by considering a thin vortical layer around the drop surface, which accounts for small irrotational shear stress at the surface \cite{lundgren1988}. However, the pressure correction for the vortical flow introduced in this formulation is applicable to weak viscous effects for droplets with Oh\textless\textless1. As for the present experimental parameters Oh$\sim$0.1, the amplitude of natural oscillations decays to zero within half cycle of applied oscillation. Moreover, the nonlinear straining term in viscous corrected formulation causes numerical instability in the oscillation pattern. Thus in the present BIM formulation viscous effects are not considered, and it is also found that the presence of natural frequency does not affect the overall frequency response of the drop surface to the driving field.} 
 
\begin{figure}[h!]
	\centering
	\includegraphics[width=0.7\linewidth]{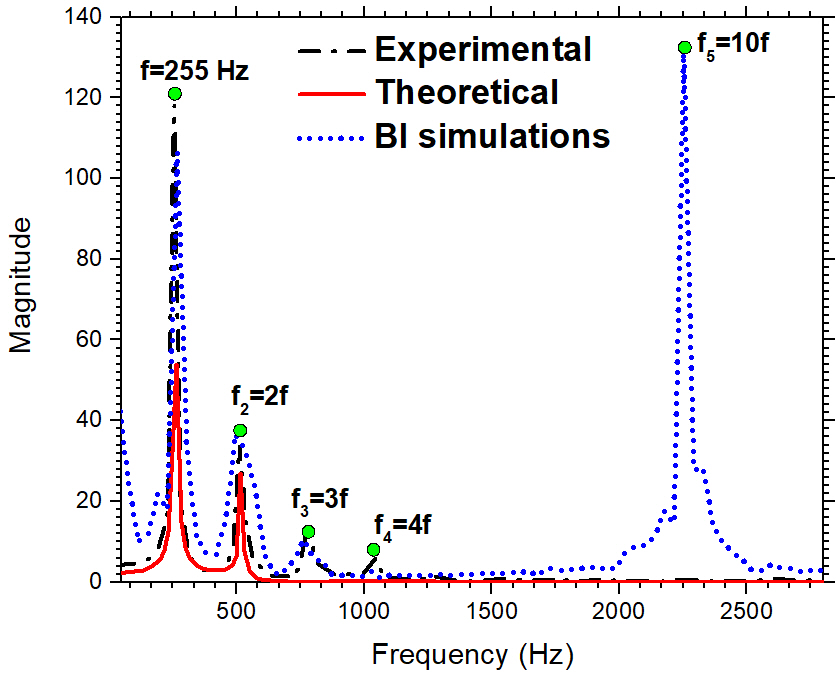}
	\caption{\textcolor{black}{FFT of experimentally observed, theoretically and numerically obtained surface dynamics for a sinusoidal waveform. The typical parameters used for theory and simulations are borrowed form experimental observations. The magnitude of theoretical FFT is scaled by the factor of 10.}} \label{sin_thoery}
\end{figure}
\begin{figure}[h!]
	\centering
	\includegraphics[width=0.7\linewidth]{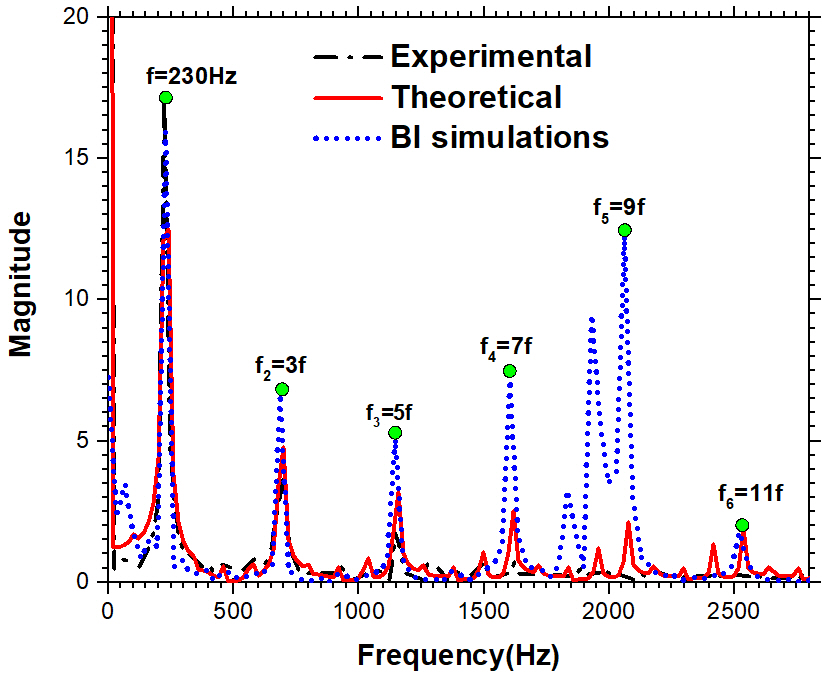}
	\caption{\textcolor{black}{FFT of experimentally observed, theoretically and numerically obtained surface dynamics for a square waveform. The typical parameters used for theory and simulations are borrowed form experimental observations. The magnitude of theoretical FFT is scaled by the factor of 5.}} \label{square_thoery_exp}
\end{figure}

\section{Validation of experimental results with theory}

Since the detailed experimental validation of surface oscillations in terms of the degree of deformation with time for a sine waveform was reported in one of our recent publications, i.e., ref.~\cite{singh2018surface}. In the present study, we have experimentally, theoretically, and numerically validated the frequency response of surface dynamics in terms of FFT analysis of surface oscillations, as shown in figure \ref{sin_thoery}.
In order to obtain the theoretical surface oscillation characteristics for a sinusoidal applied waveform, equation \ref{eqn:pot_diff2}, \ref{eqn:pot_diff3} and \ref{eqn:pot_diff4} are solved for the experimental parameters. Since the oscillations are recorded before the droplet breakup, the charge is considered as sub-Rayleigh ($X$=0.9), and the $z_{shift}$ of the droplet is kept as a fitting parameter. The $DD$ value from the theory is obtained by using the following expression,
\begin{equation}
D=\frac{3\alpha_2}{4}+\frac{\alpha_3}{2}+\frac{5\alpha_4}{16}
\end{equation}  
Similar to the experimental analysis, the embedded harmonics are resolved by performing the FFT analysis of theoretically obtained surface oscillation data with the help of the software OriginLab. It can be observed from figure \ref{sin_thoery} that there exist harmonic frequencies (in experiments) for fundamental applied frequency ($f_1$=255 Hz). The first two harmonics are expected because of the significant magnitude of the coefficient of $\zeta^2$=$\cos^2(\omega t)$ due to off-center position of the droplet, in addition to $\zeta$=$\cos(\omega t)$ due to charge on the droplet (term X$\sqrt{Ca_Q}$ in equation \ref{eqn:pot_diff2}). The resultant droplet oscillation dynamics contains both the frequencies. Similar observations are made in the theoretical oscillations, as shown in figure \ref{sin_thoery}. However, it can be also observed in the figure \ref{sin_thoery} that the experimental FFT exhibit 3$f_1$ and $4f_1$. The presence of these higher-order harmonic frequencies suggests non-linear interaction between the charges on the deformed droplet and the quadrupole field acting on these charges. Since the theory presented here is based on linear order approximation, it captures only the lower order frequencies i.e., $\omega$ and $2\omega$. The BEM simulations carried out in the potential flow limit capture these non-linear interactions and show the presence of higher harmonic frequencies along with the fundamental $f$ and $2f$ frequencies. The FFT analysis of the oscillations obtained from BEM simulations also indicates a peak at a frequency equal to $10f$. This peak corresponds to the natural inertial oscillations of the drop surface, which are not present in both the experiments and the theory as these natural oscillations are damped out by the viscous forces. \textcolor{black}{The height of the peak is a reflection of power density, so if one doubles the sampling frequency, and hence half the width of each frequency bin, it will double the amplitude of the FFT result. The good quality data in figure \ref{sin_thoery} allowed us to use smaller bin size thereby increasing the peak amplitude by a factor of 10.} 

The analysis is further extended to surface dynamics for square and ramp waveforms. In the case of a square waveform, the time-varying function $\zeta$ can be given as:
\begin{equation}
\zeta=2 \tan^{-1}\frac{[\frac{2 \pi ft}{\delta}]}{\pi},
\end{equation}
\textcolor{black}{where, $\delta(=0.01)$ is a small parameter and used for smoothing the corners of the square waveform because the sharp corner can cause the singularity.} Thus, if the value of $\delta$ is high the curve will be smooth at the corners. Figure \ref{square_thoery_exp} shows the FFT analysis of theoretical and experimental surface oscillations. It can be observed that fundamental frequency and harmonic frequencies occur at the same order. Along with harmonic frequencies, other inter-harmonic frequencies can be observed in figure \ref{square_thoery_exp}. It can be observed that, in the case of the square waveform, the oscillation pattern exhibits fundamental frequency ($f=230Hz$) and higher-order harmonic frequencies in terms of odd multipliers of the fundamental frequency i.e., $3f$, $5f$, $7f$, etc. The FFT analysis of the experimental observation shows peaks at $f$, $3f$, and $5f$, while the peaks corresponding to higher frequencies are insignificant. However, in both linear order theory and BEM simulations, the peaks at $7f$, $9f$, and $11f$ are clearly visible. Along with harmonic frequencies, other inter-harmonic frequencies can also be observed in figure \ref{square_thoery_exp}. The presence of inter-harmonic frequencies shows the complex coupling between various terms discussed in the previous section. Additionally, it is also observed that if the magnitude of $Ca_\Lambda$ and $Ca_E$ terms is higher, the FFT of theoretical surface oscillations contains even multipliers (i.e., $2f_1$, $4f_1$, $6f_1$, etc.) of the fundamental frequencies. The results of BEM simulations show higher amplitudes of the inter-harmonic frequencies near the peak of $9f$. As we have seen in the case of the sinusoidal waveform, the natural frequency due to the inertia of the system is at $10f$. Thus the presence of higher amplitudes of the peaks corresponding to inter harmonic frequencies between $7f$ and $9f$ can be attributed to the resonance between the natural and applied oscillations. \textcolor{black}{Unlike the case of a sine waveform, it is interesting to observe that the magnitude of the natural oscillation frequency peak is reduced as compared to the fundamental frequency peak. It can be hypothesized that the reduction in the peak may be due to the interaction of imposed oscillations, i.e., $7f-9f$, with the natural oscillations. Since numerical simulations are hinting an interesting aspect of non-linear oscillations, a second or higher-order theory can be helpful in result guiding, which is the future scope of the present work. }  

\begin{figure}[t]
	\centering
	\includegraphics[width=0.7\linewidth]{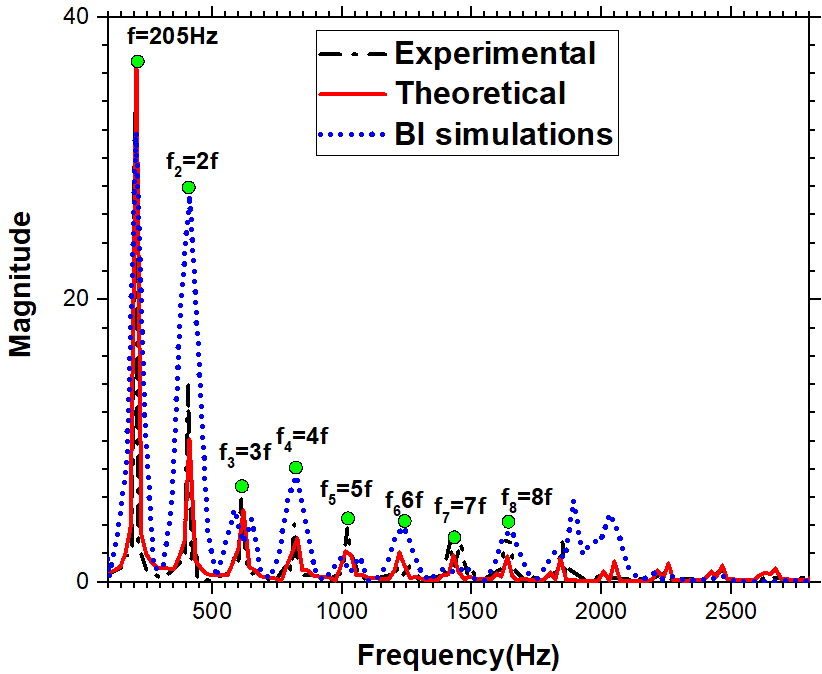}
	\caption{\textcolor{black}{FFT of experimentally observed, theoretically and numerically obtained surface dynamics for a ramp waveform. The typical parameters used for theory and simulations are borrowed form experimental observations.The magnitude of theoretical FFT is scaled by the factor of 5.}} \label{ramp_thoery_exp}
\end{figure}
Similar to the case of a square waveform the case of a ramp waveform is examined theoretically for $\zeta$, given as: 
\begin{equation}
\zeta=\frac{(t-\tau t_c[\frac{t}{\tau}])-\frac{\tau}{2}}{\frac{\tau}{2}}
\end{equation}
Where, $\tau=1/f$, $f$ is the applied frequency, $t_c$ is a defined function which accept only integer part. 
The theoretical validation of surface oscillations in the presence of ramp waveform is done by performing the FFT analysis of theoretical surface oscillations, as shown in figure \ref{ramp_thoery_exp}. Unlike the case of a square waveform, no inter-harmonic frequencies are observed in the case of ramp waveform. This is because, while in the case of a square waveform, the harmonic frequencies occur in the multiple of an odd number of the fundamental frequency, the interactions of various terms can generate even order inter-harmonic frequencies along with odd-order harmonic frequencies. On the other hand, in the case of ramp waveform, all integer harmonic frequencies are present in the applied signal itself. Like the previous case, the theoretically obtained FFT, as shown in figure \ref{ramp_thoery_exp}, is in fair agreement with the experimental observations. \textcolor{black}{In FFT analysis of the BIM results, there are many inter harmonic frequencies present between $7f$ and $9f$, and it is also interesting to observe that there is no natural oscillation frequency peak. Here we hypothesized that this is due to resonance of the natural frequency with the higher harmonic frequencies present in the applied field. The understanding of resonance in the case of an electrodynamically levitated charged droplet is not yet addressed in the literature and is the future scope of the present of work.} 

\subsection{Droplet breakup characteristics}

\begin{figure}[h!]
	\centering
	\includegraphics[width=0.7\linewidth]{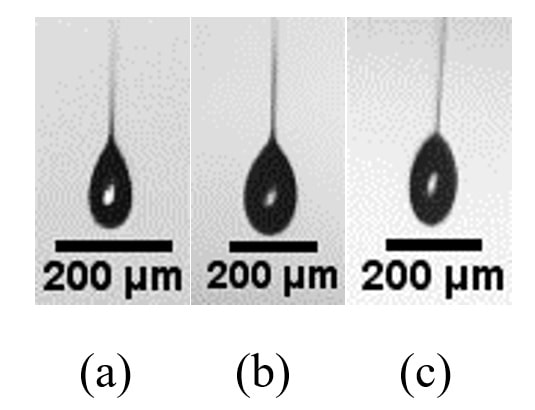}
	\caption{Breakup of levitated charged droplet in the presence of different waveform a) sine waveform b) square waveform c) ramp waveform. } \label{droplet_breakup}
\end{figure} 
To the best of our knowledge, there exist only two experimental evidence of breakup of charged droplet levitated in electrodynamic balance\cite{duft03,singh2019subcritical}. In both the studies, the droplet is levitated in the presence of a sinusoidal applied potential. In this work, for the first time, we report the droplet breakup in the presence of a non-sinusoidal waveform. \textcolor{black}{ The objective of the work is to demonstrate the robustness of the levitation process to the AC waveforms applied across the trap. The aim is also to see the response of the droplet in the sub-Rayleigh limit to different waveforms to understand the nonlinearities in the system.}  \textcolor{black}{This work demonstrates that the physics of levitation, as well as levitation-deformation coupling, can be understood by the use of different waveforms. The frequency response of the deformation is clearly seen to be governed by both the inherent frequencies in the applied waveform as well as the nonlinearities in the droplet dynamics.} Thus, the charged droplet is levitated in the presence of sine, square, and ramp waveform.
Unlike the surface oscillations of a sub-Rayleigh charged drop, for the breakup studies, we allowed the droplet to evaporate and build a Rayleigh critical charge at which droplet undergoes breakup.
The droplet breakup is recorded at one hundred thousand fps. The experimental observations of the droplet breakup in the presence of various waveform is shown in figure \ref{droplet_breakup}, where it can be observed that droplet breaks in the upward direction and in an asymmetric manner. Since the droplet is levitated in the presence of the AC field without any superimposed DC voltage to balance the mass of the drop, the droplet levitates away from the geometric center of the trap. At this off-centered location, it experiences asymmetric electrical stress, which causes an asymmetric breakup. It is also observed that the droplet breaks in the upward direction due to the high initial $P_2$ perturbation and higher curvature (+ve $P_3$) in the upward direction. The magnitude of asymmetry and droplet DD depends upon the $z_{shift}$ and $\Lambda$. As reported by \citet{singh2019lang}, the droplet breaks with more asymmetry at a higher value of $z_{shift}$ even for a fixed value of $\Lambda$. The thickness of the jet depends on the droplet diameter (or $z_{shift}$) and the applied voltage. The detailed explanations of such droplet breakup characteristics are discussed by \citet{singh2019lang}. The important part to note here is that there is no significant difference in the breakup mode in the presence of different waveforms. 

\section{Conclusions:} 
\textcolor{black}{The study aims to understand various aspects of droplet oscillations of a levitated charged drop in to ultimately make droplet levitation as tool for measuring interfacial properties. In that pursuit, we have levitated the droplet in presence of various waveforms and it is observed that the non-linearity is higher in case of sine waveform while the droplet oscillations follow the applied signal in case of square and ramp waveforms.} The surface oscillation characteristics are investigated by performing FFT analysis and it is found that the FFT is an appropriate characterizing tool for identification of the presence of different frequencies in the droplet deformation data. The surface oscillation behavior is also compared with the viscous-potential flow theory and a fair agreement is observed. The droplet breakup characteristics in the presence of sinusoidal and non-sinusoidal waveform are shown and it is observed that the application of a different waveform does not alter the breakup characteristics. The extension of linear stability analysis to higher-order analysis can make the system suitable for a contact-free and accurate surface tension measurement device. The work demonstrates that the mechanism of droplet oscillations and breakup is robust and is admitted as long as there is a time periodic driving potential applied to the trap. 

\subsection*{Data availability statement}
The data that support the findings of this study are available from the corresponding author upon reasonable request.

\providecommand{\noopsort}[1]{}\providecommand{\singleletter}[1]{#1}
%


\end{document}